\documentclass[10pt,journal,compsoc]{IEEEtran}
\usepackage{color}
\usepackage{xcolor}
\definecolor{myblue}{RGB}{179,205,227}
\definecolor{myred}{RGB}{251,180,174}
\definecolor{mygreen}{RGB}{204,235,197}
\definecolor{textred}{RGB}{228,26,28}
\definecolor{textblue}{RGB}{55,126,184}
\definecolor{textgreen}{RGB}{77,175,74}
\usepackage{url}
\usepackage{enumitem}
\usepackage{multirow}

\definecolor{text8}{RGB}{67, 109, 158}
\definecolor{text9}{RGB}{243, 122, 23}
\definecolor{text10}{RGB}{224,77,76}
\definecolor{text4}{RGB}{103,174,169}
\definecolor{text5}{RGB}{74,152,66}
\definecolor{text6}{RGB}{235,195,52}
\definecolor{text7}{RGB}{169,110,152}
\definecolor{text1}{RGB}{255,147,156}
\definecolor{text2}{RGB}{147,106,82}
\definecolor{text3}{RGB}{177,167,162}

\newcommand*{\green}{\textcolor{textgreen}}

\usepackage{tikz}
\newcommand*\circled[1]{\tikz[baseline=(char.base)]{
            \node[shape=circle,draw,inner sep=1pt] (char) {#1};}}

\usepackage{graphicx}
\usepackage{caption}
\usepackage{ulem}

\usepackage{algorithm}
\usepackage[noend]{algpseudocode}

\usepackage{makecell}

\usepackage{amsmath}

\ifCLASSOPTIONcompsoc
  \usepackage[nocompress]{cite}
\else
  \usepackage{cite}
\fi
\ifCLASSINFOpdf
\else
\fi
\newcommand{\sure}{\textsc{SuRE}}

\begin{document}
%
\title{Visual Exploration of Machine Learning Model Behavior with Hierarchical Surrogate Rule Sets}





%
%
%
%

\author{Jun Yuan, Brian Barr, Kyle Overton, and Enrico Bertini
\IEEEcompsocitemizethanks{\IEEEcompsocthanksitem Jun Yuan and Enrico Bertini are with New York University.\protect\\
E-mail: \{junyuan, enrico.bertini\}@nyu.edu
\protect\\
\IEEEcompsocthanksitem Brian Barr and Kyle Overton are with Capital One. \protect\\
E-mail: \{brian.barr, kyle.overton\}@capitalone.com

}

\thanks{Manuscript received January, 2022.}
}

%
%

\markboth{}%
{Visualizing Rule Sets}
%



\IEEEtitleabstractindextext{%
\begin{abstract}
    One of the potential solutions for model interpretation is to train a surrogate model: a more transparent model that approximates the behavior of the model to be explained. Typically, \textit{classification rules} or \textit{decision trees} are used due to the intelligibility of their logic-based expressions. However, decision trees can grow too deep and rule sets can become too large to approximate a complex model. Unlike paths on a decision tree that must share ancestor nodes (conditions), rules are more flexible. However, the unstructured visual representation of rules makes it hard to make inferences across rules. To address these issues, we present a workflow that includes novel algorithmic and interactive solutions. First, we present \textit{H}ierarchical \textit{S}urrogate \textit{R}ules (HSR), an algorithm that generates hierarchical rules based on user-defined parameters. We also contribute {\sure}, a visual analytics (VA) system that integrates HSR and interactive surrogate rule visualizations. Particularly, we present a novel feature-aligned tree to overcome the shortcomings of existing rule visualizations. We evaluate the algorithm in terms of parameter sensitivity, time performance, and comparison with surrogate decision trees and find that it scales reasonably well and outperforms decision trees in many respects. We also evaluate the visualization and the VA system by a usability study with 24 volunteers and an observational study with 7 domain experts. Our investigation shows that the participants can use feature-aligned trees to perform non-trivial tasks with very high accuracy. We also discuss many interesting observations that can be useful for future research on designing effective rule-based VA systems.

\end{abstract}

\begin{IEEEkeywords}
visualization, rule set, surrogate model, model understanding.
\end{IEEEkeywords}}

\maketitle

\IEEEdisplaynontitleabstractindextext

%
\IEEEpeerreviewmaketitle

\IEEEraisesectionheading{\section{Introduction}\label{sec:introduction}}

%
%
%
%

In this paper, we describe a visual analytics method and a system to help data scientists and model developers explore and analyze the behavior of machine learning classifiers. The approach we propose explores the idea of inspecting an existing model by (1) generating hierarchical rules that \textit{describe} the model's decision space and (2) visualizing the results with an interactive hierarchical visualization that depicts the extracted rules.


Explaining model behaviors has become one of the most important needs for AI/ML systems. Given the often-opaque logic and structure of models and their enormous complexity (e.g., with the employment of ensemble and neural network models), it has become necessary to develop methods to help experts navigate their decision space.
While common aggregate statistics continue to be the backbone of validation, being able to understand how the model makes decisions and validate the decision-making logic has become equally relevant; especially in systems where a model making decisions for wrong reasons can lead to drastic consequences (e.g., health care, security, law enforcement). 

Model inspection has been addressed from many different angles. A common distinction of these methods is between solutions that look at model behavior \textit{locally}, for one single instance of interest at a time; and solutions that aim at providing a description of how the model behaves \textit{globally}, either on the entire input space or in meaningful data subgroups. Our work aims at supporting experts to deal with this latter problem of how to explain a model globally, where typically experts want to (1) verify whether model decisions make sense; (2) identify the errors made by the model in subgroups; and (3) verify how the model behaves with specific subgroups of interest. Within this area of investigation, researchers can distinguish between solutions for models that are inherently \textit{transparent}, that is, models with structures amenable to human inspection, and models that are \textit{opaque}, that is, models with structures that are too complex for direct human inspection. Since this last class of models does not directly expose an intelligible structure to their users, methods are necessary to first extract information out of them to enable inspection.

In this work, we focus on this global model understanding problem and, more specifically, on a class of solutions called ``surrogate models''. The basic idea behind surrogate models is to train a transparent model using the output of the (opaque) model one wants to investigate. By doing this, the surrogate simulates the decision space of the model of interest and provides a description of its behavior (e.g., model predictions and errors in a data subgroup). For instance, under this paradigm, an ``opaque'' classifier trained using a neural network could be simulated by a ``transparent'' decision tree trained using the labels generated by the neural network as an input to its training process. 

Due to the intelligibility of logic-based models, surrogates are typically generated using either \textit{classification rules} or \textit{decision trees}. However, these existing approaches have several limitations. Decisions trees are often not flexible enough to simulate the decisions of more complex models. Also, due to their rigid binary tree structure, decision trees need to grow deep and large to simulate complex models accurately; thus, nullifying the original intent of creating an intelligible surrogate. Rules are more flexible than trees because each rule can be generated independently from the other rules, however, rules can also generate very large outputs and their lack of an organizing structure makes it hard to make inferences that involve multiple rules. For instance, it is hard to identify if all rules that share a common condition tend to have a similar outcome, or two similar rules differ only from one small condition that has a strong impact on the outcome (i.e., the two rules lead to different predictions despite being very similar).

In order to solve these problems, we propose a new approach that integrates the benefits of trees and rules while trying minimizing their shortcomings. The approach consists in the generation of ``Hierarchical Surrogate Rules'' (HSR): rule sets that are organized around a hierarchical structure. These hierarchical sets have the flexibility necessary to capture complex relationships; thus, making them more powerful than single decisions trees. At the same time, these sets have the structure necessary to enable inferences across rules; therefore making them more powerful than traditional rule sets. Similar to other rule approaches, end-users can change HSR parameters according to desired properties. In particular, users can manipulate four properties that are relevant for model exploration. These properties are (1) \textit{rule fidelity} (how well a rule simulates the model); (2) \textit{rule coverage} (what proportion of the model a rule can simulate); (3) \textit{rule length} (how interpretable a rule is); and (4) \textit{feature granularity} (the level of details presented to users).

In addition to producing hierarchical rules, we also address the problem of visualizing the rules. Traditional rule systems visualize the logic statements using textual representations, which limits interpretation in a few ways. First, existing representations do not provide an easy way to identify rules that share a common set of features or conditions. However, being able to do so is crucial to understand the effect of individual features as well as their combinations beyond the specifics of individual rules (the problem of drawing inferences across rules that we mentioned above). Second, each rule and its individual components have relevant statistics that are useful to compare visually. For instance, knowing how many data points are covered by a rule and what their associated labels are is necessary to identify subgroups of interest. Considering these limitations, we propose a rule visualization based on a feature-aligned tree structure, which organizes the rules extracted by our algorithms and enables interactive exploration. \looseness=-1

In order to validate and explore these ideas, we developed a visual analytics system that integrates the rule generation functionality and the interactive visualization we mentioned above. The validation is organized around three main activities. First, we provide algorithm benchmarking to characterize the behavior of the rule generation procedure according to different parameter settings. Within this context, we also compare the output generated by traditional decision trees and our proposed solution to elucidate the benefits of using our type of rules. Second, we describe the results of a usability study we ran to verify the readability of our visual representation. As you will see in Section \ref{sec:eval}, novice users can read the generated rules with minimal mistakes despite the novelty of the representation. Finally, to evaluate the whole interactive visual analytics system, we describe the results of a qualitative study we conducted by observing and interviewing a group of experts from a private banking company who used our tools to explore a financial data set of interest. The results of the observational study provide useful insights on the usefulness of our approach as well as interesting design inspirations.

In summary, in this research we contribute the  (1) a novel rule generation and exploration procedure to explore the decision space of a black-box model; (2) a novel feature-aligned tree visualization of rules that permits users to draw useful inferences across the rules of a rule set; (3) the design of an interactive visual analytics tool that integrates interactive rule parameterization, generation, and visualization; (4) validation studies that provide useful insights on the proposed method and on potential future directions for this area of research.

The rest of the paper is organized as follows. Section 2 presents the related work. Section 3 provides an overview of the workflow of generating and visualizing rules to explain model behavior, as well as the constraints and objectives we take into consideration for rule generation. Section 4 describes the algorithm of rule generation and then in Section 5 we present two experiments to evaluate the rule generation method. Section 6 demonstrates the design rationale of hierarchical rule visualization and interactive functions included in the entire workflow for rule exploration. Section 7 evaluates the rule visualization and the whole interactive workflow based on human-centered studies. The lessons learned, limitations and future research directions are then discussed in Section 8.
\section{Related Work}

\subsection{Surrogate Models}
A surrogate model aims at describing the original model's decisions using a structure that is more interpretable than the one used in the model it tries to simulate. Instead of accessing and showing the original model's internal structure, the method of surrogate modeling treats the original model as a black box. By showing the logic of the surrogate model, model developers and domain experts can gain an understanding of the behavior of the underlying black-box model. Models used as global surrogates are therefore usually interpretable models such as linear models~\cite{hastie1990generalized}, decision trees~\cite{safavian1991survey}, or decision rules ~\cite{frank1998generating} and generalized additive models (GAMs)~\cite{hastie2017generalized,hohman2019gamut}, which use shape functions instead of weights in linear models for each feature. Because the \textit{if-then} structure of a rule semantically resembles the decision-making process and is common in our daily lives~\cite{molnar2019}, rules are often used as model explanations to show to those who do not have a strong data science background. Therefore in this work, we focus on the generation and presentation of decision rules as global surrogate. 

To generate rules to approximate a black-box model, people can directly extract rules based on the black-box model's input and output using rule-based models such as Apriori~\cite{agrawal1994fast}, Tertius~\cite{flach2001confirmation}, Closet~\cite{pei2000closet}, etc. The surrogate rules method has been applied widely in recent research. Lakkaraju \textit{et. al} introduced MUSE~\cite{lakkaraju2019faithful} which uses rules to explain model behaviors in user-defined subspaces. More recently, LoRMIkA~\cite{rajapaksha2020lormika} extracts k-optimal association rules to explain the model predictions for classification data sets. Moradi and Samwald~\cite{moradi2021post} also apply association rules to approximate a target black box. Tree-based models can also be a source of surrogate rules by transforming tree paths into plain \textit{if-then} rules~\cite{pal2001fuzzy, quinlan1987generating}. In recent years, more and more work has been proposed to explain complex models using trees. For example, TREPAN~\cite{craven1996extracting}, RxREN~\cite{augasta2012reverse} and DeepRed~\cite{zilke2016deepred} extract tree structures from trained deep neural networks. Bastani ~\textit{et al.}~\cite{bastani2017interpreting} introduced a model-agnostic method to induce a decision tree from a black-box model. Thiagarajan \textit{et al}.\cite{thiagarajan2016treeview} applied a similar strategy to extract and visualize a surrogate decision tree to explain model behavior of a deep neutral network. 
The hierarchical structure of rules extracted from trees typically contains information about the order and relevance of the features; which is useful to understand how well these features split the data space into different classes. For this reason, we first train a tree-based surrogate model and then adapt the method of transforming tree paths into surrogate rules to explain the model behavior in this work.

Although a surrogate model tries to simulate how the black-box model makes decisions, it is important to realize that a surrogate model is distinct from the actual black-box model. To understand how much the explanation is faithful to the model to be explained, it is important to evaluate the \textit{fidelity} of the surrogate model~\cite{velmurugan2021evaluating}. One way to test the fidelity is the R-squared measure~\cite{cameron1997r}, which is widely used for regression models. Recent work ~\cite{messalas2019model,velmurugan2021evaluating} explored ways of measuring \textit{internal fidelity}, which measures the similarity of feature attribution values (e.g., SHAP~\cite{lundberg2017unified}) from the surrogate model and the black-box model. These measurements are used to evaluate the fidelity of the surrogate model in terms of the whole data space. In our work we measure the fidelity of the surrogate rules at a subset level because each rule is an independent classifier and can only describe a subset of instances in the data space. 

Human-interpretability is also considered when generating and evaluating a surrogate model. In practice, model developers and domain experts can understand a black-box model by understanding its surrogate; which highlights the importance of the interpretability of a surrogate model. In recent work, researchers found that model interpretability can be measured and manipulated~\cite{poursabzi2021manipulating} based on user studies with a regression model and different explanations. In terms of rule generation, researchers identified a few properties of interpretable rules in the work of interpretable decision sets\cite{lakkaraju2016interpretable} and Bayesian rule sets~\cite{wang2017bayesian}, such as the number of rules and the number of conditions in a rule. Lage \textit{et. al}~\cite{lage2019evaluation} also introduced three types of model complexity based on literature from cognitive psychology: explanation size, cognitive chunks, and repeated terms. Inspired by these work, we introduce a few constraints and objectives in the surrogate rule generation step to ensure the surrogate rules are interpretable while maintaining fidelity high. \looseness=-1

\begin{figure*}
 \centering 
 \includegraphics[width=.9\textwidth]{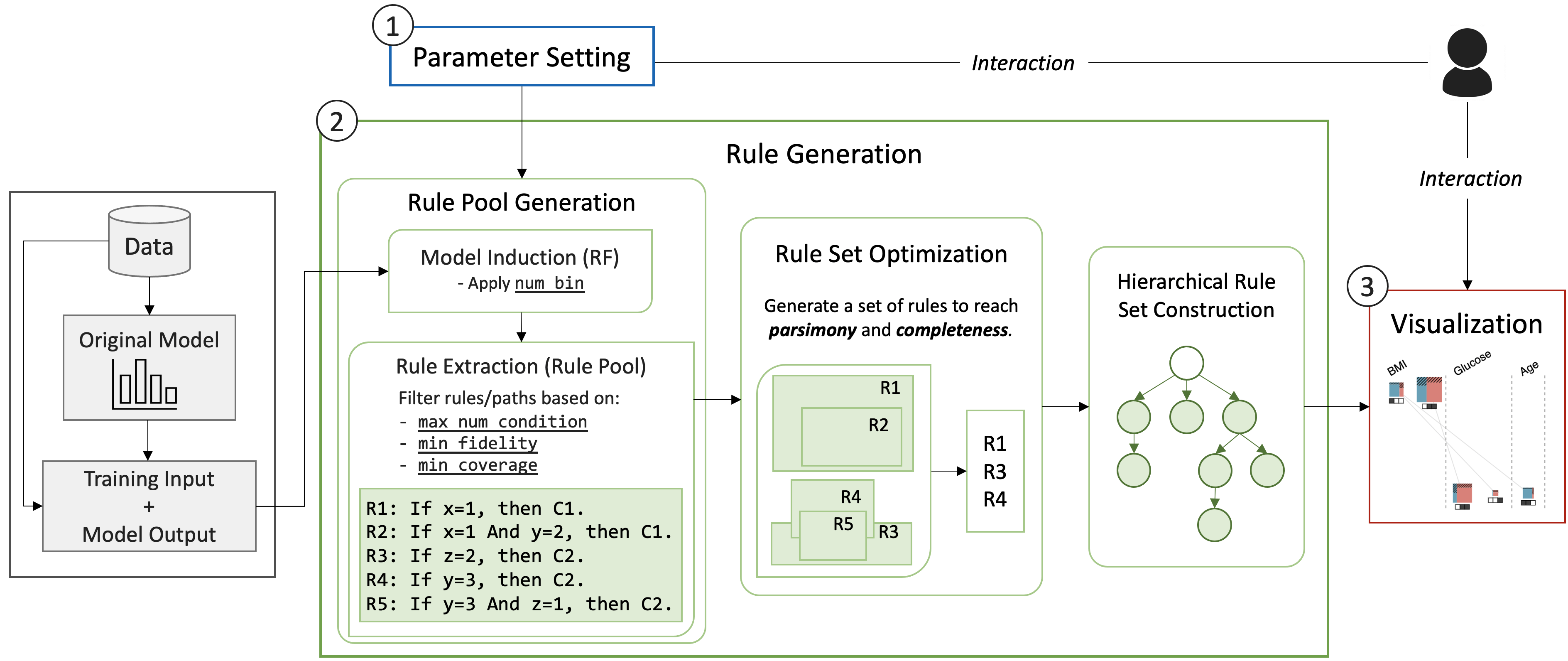}
 \caption{The proposed workflow consists of three main components: (1) parameter setting based on user's requirements, (2) rule generation to achieve the constraints and objectives, as well as the hierarchical structure extraction for hierarchical visualization; (3) visualization of the hierarchical rule set. Users are involved in this workflow by setting parameters and interacting with the visualization directly.}
 \label{fig:pipeline}
\end{figure*}

\subsection{Visual Representations of Rules and Decision Trees}

While in recent years research on rule-based models has grown substantially, the problem of how to visualize the rules has gain COMPARATIVELY little attention. Decision rules are mostly presented as plain textual  \textit{if-then} statements, independently of whether the rules are used as model explanations ~\cite{lakkaraju2017interpretable,lakkaraju2019faithful,guidotti2018local,ribeiro2018anchors} or are used directly as interpretable predictive models ~\cite{lakkaraju2016interpretable,wang2017bayesian,wang2015falling}. For rules extracted from a decision tree~\cite{safavian1991survey}, tree structures have been studied widely in the visualization research community (a gallery of tree visualization techniques can be found at treevis.net~\cite{schulz2011treevis}) but only with little focus to structures designed to explore model behavior. To assist domain experts to construct and analyze decision trees, Elzen \textit{et. al}\cite{van2011baobabview} introduced \textit{BaobabView}, which visualizes model prediction behaviors in the nodes of a decision tree.
Recent work from REMIX~\cite{zarlenga2021efficient} visualizes a set of rules using a single n-ary tree. However, with a traditional tree structure, a feature of interest appears in different nodes distributed across the tree. This makes it hard to analyze rules or features along the whole tree. In our work, we propose a feature-aligned tree visualization to assist rule understanding and analysis across multiple rules. Moreover, if rules come from large sets of trees such as random forest, it is challenging to visualize the content of all the trees in a forest. We also introduce an algorithm in Section~\ref{section:algo} to select eligible rules to present to users.
\looseness=-1

To improve the interpretability of rules, we believe proper rule visualization can help. Recent research reveals that the understanding of rules is relevant to several visual factors, including feature alignment, condition encoding, ordering of rules and features~\cite{yuan2021exploration}. In the same paper, the authors also conducted a controlled study on a set of rules in different visual representations showing that feature alignment has a strong impact on how efficiently readers can process the rules while having minimal impact on accuracy. In another paper, Lakkaraju et al.~\cite{lakkaraju2016interpretable} compared Bayesian rule lists~\cite{letham2015interpretable} with \textit{decision rule sets} and found that the \texttt{if-then} structure of rule sets, as opposed to the \texttt{if-then-else} structure of rule lists, improves the accuracy and efficiency of rule interpretation. Studies also exist on comparing hierarchical versus flat (list) organizations of rules. Subramanian et al.~\cite{subramanian1992comparison} investigated how well people can simulate investment strategies (decision making) showing rules either in a tabular or hierarchical format and found that trees are better for decision support than a hierarchical table for interpreting the conditional logic of rules. Another similar study compared decision trees, structured \texttt{if-then-else} rule lists, and tables~\cite{vessey1986structured} and found that structured text outperformed decision trees and tables when the goal was to identify specific conditions to match an action. Huysmans et al.~\cite{huysmans2011empirical} evaluated three different representations of rules: decision tables, decision trees, and lists. The results showed that tables led to the best performance. Based on these results, in this work, we propose a novel visual representation for rule sets which integrates the benefits of both hierarchical structure and tabular (aligned) layout with proper encoding of rule content and properties. \looseness=-1

In recent years, some visual analytics solutions have been developed to explore models through trees and rules. 
Most related to our work, \textit{RuleMatrix}~\cite{ming2019rulematrix} proposes a tabular layout to visualize the logic of a list of rules. The conditions of each rule are organized in a tabular layout where the conditions that use the same features are aligned in the same column; which makes it easier to compare data distribution of the same feature in different rules. Some recent work, such as Surrogate Decision Trees~\cite{di2019surrogate} and Explainable Matrix~\cite{neto2020explainable}, also apply a flat layout to take advantage of feature alignment for visualizing the rules. However, the rules visualized in these works come from one or more trees which have a hierarchical structure. Showing them in the flat tabular layout they use neglects the important feature/condition ordering information embedded in the hierarchical structure. More recently, Vladimir \textit{et. al} ~\cite{estivill2020human} applied parallel coordinates to visualize the splits on a decision tree. However, such representation has the same problem of not maintaining the hierarchy of a tree. Another related work is \textit{iForest}~\cite{zhao2019iforest}, which provides an interface to analyze specific decision paths generated by the Random Forests (RFs) method. This solution has the advantage of showing the hierarchical structure generated by RFs, but by doing that it neglects to visualize the logic of rules and find useful logical relationships between rules that come from separate trees. In this paper, our work aims at finding a synergy between all these approaches. We developed a solution that visualizes the hierarchical structure of rules while retaining a layout that provides information about the logic of the rules. \looseness=-1



\section{Workflow, Constraints and Objectives}
\label{section:pipeline}

The workflow we propose consists of three main steps and associated components, which we exemplify in Figure~\ref{fig:pipeline} (see numbered steps): \textbf{(1) parameter setting:} the user sets the desired parameters for the generation of rules. The parameters provide information and constraints to the rule generation algorithm in terms of desired faithfulness and complexity constraints. Below we describe in more details what these parameters are and how they are computed; \textbf{(2) rule generation:} the system takes as an input the data, the model and the parameters set by the user (or set by default), and then produces a set of hierarchical rules. In Section~\ref{section:algo} below, we describe how the rules are structured and how they are generated by the algorithmic procedure we propose; \textbf{(3) rule visualization and exploration:} the system visualizes the rules using innovative data visualization techniques, including the \textit{Feature-Aligned Tree}, a layered network visualization that depicts the hierarchical structure of the rules and helps orient the user towards rules and patterns of interest. In Section~\ref{section:vis}, we describe the design of the visualizations, providing the rationale behind the design and demonstrating the utility of these visual techniques.

The rule generation procedure takes as an input training data and model output and uses them to build a set of \textit{descriptive} rules that simulate the original model. Each rule has several conditions followed by an outcome that represents the rule's prediction. More precisely, the output of the procedure is a hierarchical list of rules where each rule has the following format: \texttt{IF $c_1\wedge c_2\wedge ...$, THEN $class_1$.} where $c_i$ is the $i$th condition in the rule and $class_j$ is the consistent prediction made by the black-box model when all the conditions are matched. The if-part of a rule is typically called \textit{antecedent} and the then-part \textit{consequent}. Each condition is a logical predicate specifying the value of a feature in the data set (e.g., Age $> 50$).

Each surrogate rule effectively describes a specific subset of the input space (the instances that satisfy the conditions in the antecedent part of the rule) and provides a model prediction for it as the rule consequent (e.g., \textit{non-diabetic} for a diabetes prediction problem). 

Next, we define the properties of surrogate rule sets that we want to manipulate and optimize for in order to reach two main purposes: low complexity and high faithfulness. At the level of individual rules, we want the rules to be faithful and interpretable, whereas at the level of the whole set, we want the set of rules to cover as much of the model behavior as possible while remaining as simple as possible in terms of total number of rules. In the rest of the paper, we will refer to \textit{constraints} that are applied at the level of \textit{individual} rules and that will be set directly by the user.  Then, we will refer to \textit{objectives} that are applied at the level of the whole \textit{set} of rules and will be optimized by the algorithmic procedures we describe in the next section. 


The \textbf{faithfulness} at the level of an individual rule is described by \textit{fidelity} and \textit{coverage}. A rule simulates the original model faithfully if it has high \textit{fidelity}, that is, the consequent (model prediction given by the rule) is the same as the actual model predictions to all (or most of) the data instances the rule describes. Each rule is also evaluated by \textit{coverage}, which is the number of data instances the rule covers with its antecedent part. In this research, we consider a rule as a faithful description of a set of data instances if it reaches high fidelity and high coverage. 

Since the goal of surrogate rules is to enable human inspection, another relevant goal is to reduce \textbf{complexity} as much as possible, while keeping fidelity and coverage high. Complexity can be measured in many ways at the level of individual rules. We focus on two main complexity parameters in this work: \textit{rule length}, which is the number of conditions in a rule and \textit{feature granularity}, which is the number of values each feature can refer to. A rule is considered simpler if it has a fewer number of conditions and a fewer number of values or ranges the features in the rule can refer to. Given a condition of "Skin Thickness $< 31$", users need to think about whether this metric is actually low or high relative to other patients. Therefore, we enable users to control the feature granularity as the number of bins assigned to each feature. For example, instead of using many different numeric ranges, we allow users to use ordinal descriptors such as \textit{low}, \textit{medium}, and \textit{high} when the number of bins is 3. Typically, the lower the \textit{feature granularity}, the less complex the rule is.

While the constraints we just described work at the level of individual rules, the algorithmic procedure still has to optimize the results globally according to a number of additional objectives. In this paper, we have identified two main \textit{objectives} that we want the procedure to pursue at the level of a rule set, rather than an individual rule. 
The first one is \textbf{completeness}, that is, the rules should cover as many instances as possible to make the surrogate rule set faithful enough to describe the overall model behavior. When we look at the whole set of rules generated by the procedure we can compute the \textit{rule set coverage}, which is the proportion of the whole data set covered by the generated set of rules. The coverage of the rule set gives us a sense of how much of the original model can be described by the rule set. For this reason, one of the major objectives of this paper is to generate a rule set that covers as much as possible of the data space. The second objective is \textbf{parsimony}, that is, the \textit{number of rules} in a rule set should be as small as possible. When the number of surrogate rules is too high, it is hard for people to analyze all of the rules and draw useful inferences from the comparison of rules; especially when there are hundreds or even thousands of rules. Therefore, the rule set generation method needs to produce as few as possible rules to reduce the rule set level complexity. \looseness=-1

It's important to notice how the purposes of high faithfulness and low complexity may be in opposition. For instance, to cover as much as possible of the data set it is necessary to create more rules. Similarly, if we want to have high-fidelity rules, we typically have to produce rules with more conditions. At the same time if each rule has more conditions it also means it will cover fewer data instances, which in turn, means more rules will be needed to achieve a high level of coverage. 

Since we cannot know in advance what trade-offs are more appropriate for a given problem, model, data and user needs, it is important to provide flexibility and parameterization so that the algorithmic procedure that produces the rule set can find an optimized solution taking the parameters set by the user into account.


Given a trained model $F$, and a surrogate rule set $R$ that approximates the behavior of $F$ using a data set $\mathcal{D}$, we define a set of metrics that characterize individual rules. For a given rule $r \in R$ covering the data subset $\mathcal{D'} \subseteq \mathcal{D}$ (that is, the subset of the data for which the antecedent of the rule is true) we define the following metrics.

\begin{itemize}
    \item \textbf{Rule fidelity}: we define $fid(r)$ as the percentage of instances $x$ within $\mathcal{D'}$ for which $F(x) = r(x)$, where $F(x)$ is the prediction of the model on $x$ and $r(x)$ is the prediction of rule $r$ on $x$. To include rules with high enough fidelity, we allow users to set \texttt{min\_fidelity} as a minimum fidelity for each rule.
    
    \item \textbf{Rule coverage}: we define $cov(r)$ as $|\mathcal{D'}|/|\mathcal{D}|$, the number of instances covered by the rule over the total number of instances in the data set. To include rules that cover enough instances, we allow users to set \texttt{min\_coverage} as a minimum coverage threshold for each rule. 
    \looseness=-1
    
    \item \textbf{Rule length}: we define $len(r)$ as the number of conditions contained in the antecedent of the rule, which can be controlled by the parameter \texttt{max\_num\_condition}.
    
    \item \textbf{Feature granularity}: we define $num\_bin$ as the number of bins assigned to the values of each feature. Users can control it by setting the value of \texttt{num\_bin}.
\end{itemize}




In the following section we describe the algorithmic procedure we devised to create rules based on the constraints and then optimize the rule set to reach the objectives we described above.

\section{Hierarchical Surrogate Rule Sets}
\label{section:algo}
The algorithmic procedure we proposed to achieve the goals presented in the previous section is based on three main phases, a generative phase we call (1) \textit{rule pool generation}, an optimization phase we call (2) \textit{rule set optimization} and a structuring phase called (3) \textit{hierarchical rule set construction}. The first phase aims at producing an initial set of surrogate rules that satisfies the constraints set by the user-selected parameters. The second phase aims at creating an optimal set of rules according to the objective of parsimony and completeness. The third phase aims at organizing the extracted rules into a hierarchical structure.

\subsection{Rule Pool Generation}

The rule generation procedure includes two main steps: (1) \textit{model induction}, to build a surrogate random forest and (2) \textit{rule extraction}, to extract rules that satisfy the constraints set by the user from the random forest.

To construct a large pool of qualified rules we feed a \textit{random forest} with the labels generated by the original model rather than the labels included in the original data. Random forest is a good candidate for our purposes because the bootstrap sampling of the data and the random selection of the features it employs ensures that many feature combinations and data subspaces are considered in the process. This is important to ensure the model can simulate the original model with sufficient fidelity and that a large set of candidate rules is generated. The random forest algorithm requires a number of trees to create when building the model. We use a default value of 100, but this also is a parameter that can be manipulated by the user if necessary. 
\looseness=-1

To control the feature granularity, we first transform continuous data into ordinal values. In this work, percentiles are used to control the thresholds that generate the original bins from quantitative data. 
By default, the bin number is set to three (low, medium and high) and we let the user decide the granularity they prefer by changing \texttt{num\_bin}.
\looseness=-1



\begin{figure}
 \centering 
 \includegraphics[width=\columnwidth]{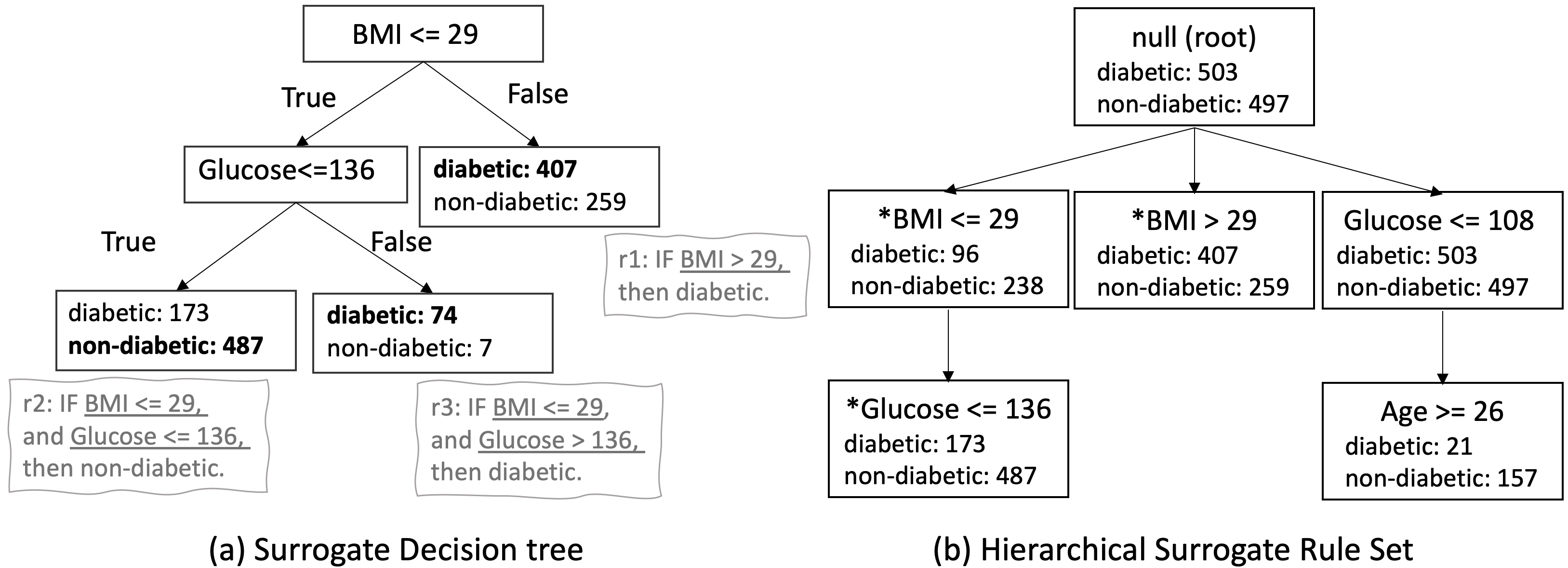}
 \caption{We can extract three decision rules from (a) surrogate decision tree (r1, r2, r3). The hierarchical structure of rules from multiple trees can be shown as (b) hierarchical surrogate rule set, where the paths with a star symbol (*) are from the decision tree on the left (r1, r2), and the other rule path comes from another decision tree in the forest.}
 \label{fig:rule_tree}
\end{figure}

Once the random forest is generated, the next step is to extract rules that satisfy the parameters/constraints we described in Section~\ref{section:pipeline} out of the trees generated by the random forest. Each node of each tree is equivalent to a rule where the conditions found in path leading to that node represent the conditions of the rule. For this reason we can use these constraints to prune paths in the trees.


The algorithm (sketched in Algorithm~\ref{algo:tree_pruning} using pseudo-code) uses the constraints on \textit{minimum coverage}, \textit{minimum fidelity} and \textit{maximum rule length} to search each decision tree in the forest for nodes (rules) that satisfy the constraints. More precisely, the algorithm visits each tree starting from the root node in a depth-first fashion. Every time it finds a node that satisfies all the three constraints, the algorithm returns the node (rule) and disregards the subtree below that node.
\looseness=-1

It is important to note that rule coverage decreases and rule length increases monotonically with the depth of the tree; therefore the deeper down the tree, the more likely the algorithm is to find a node that does not satisfy the coverage and length constraints. Conversely, fidelity can typically only be achieved moving down deeper in the tree structure, where nodes become purer (according to how decision trees are built to optimize node purity; which in our context corresponds to fidelity).

Note that the actual implementation of the algorithm we present in Algorithm~\ref{algo:tree_pruning} returns the first node it finds that satisfies all the constraints at once, disregarding nodes in its subtree that could potentially also satisfy all the constraints. By doing that we privilege smaller rule length and higher rule coverage as a guiding principle, rather than potentially higher fidelity. In other words, despite that we could find nodes with even higher fidelity by exploring the subtree further, we prefer to return the node that achieves the shortest length and the largest possible coverage. 

\begin{algorithm}
    \caption{Extract Rules from Pruned Trees}
    \label{algo:tree_pruning}
    \textbf{Input:} The trained random forest, which is equivalent to a set of trained decision trees: $T$.\\
    \textbf{Output:} A rule pool $R$ that contains rules from pruned trees.
    \begin{flushleft}
    \begin{algorithmic}[1]
        \State $R \gets \emptyset $ \Comment{initialize an empty rule pool.}
        \For{$t \in T$}
            \State $r \gets null$ \Comment{initialize rule $r$.}
            \State $root \gets 0$ \Comment{the node id of a root is 0.}
            \State $\Call{searchNode}{root, r, t}$
            
        \EndFor
        
      \\ 
      
        \Function{searchNode}{$node\_id$, $r$, $t$} 
        \If{
                $cov(r) \geq min\_coverage$\\
                \hskip\algorithmicindent 
                $\And 1 \leq len(r) \leq max\_num\_condition$ \\
                \hskip\algorithmicindent $\And fid(r) \geq min\_fidelity$}
                    \State $R = R \cup r$  
                    \Comment{add rule $r$ to rule pool then stop.}
                    \State \Return
            \EndIf
            \If{$node\_id$ is leaf}
                \State \Return
            \EndIf
            \State  $r' = r \wedge t[node\_id].condition$ \Comment{update rule.}
            \State  $searchNode(t[node\_id].left\_child, r', t)$
            \State  $searchNode(t[node\_id].right\_child, r', t)$
        \EndFunction
    \end{algorithmic} 
    \end{flushleft}
    \vspace{-0.2cm}
\end{algorithm}

\subsection{Rule Set Optimization}
After the rule pool generation phase, we typically end up with a large number of rules. Many of these rules can be redundant in terms of which subsets of the data space they cover. The second phase of this algorithmic procedure therefore takes the output generated by this procedure and optimizes the set according to the objectives of parsimony and completeness outlined in Section~\ref{section:pipeline}.

The goal of the rule optimization step is to find the \textit{smallest} set of rules that covers as much of the data set $\mathcal{D}$ as possible. To better understand the problem, we define it in a formal way. Given a set of pre-mined rules $R=\{r_1, r_2, ..., r_m\}$ we obtain from the previous steps, each rule $r_i$ can cover a subset of instances in the data set. Here $|R|=m$ is the number of rules. A set of subsets is defined as $S=\{s_1, s_2, ..., s_m\}$, where $s_i$ is the subset of instances that match the antecedent of $r_i$. In most cases, $\cup_{i=1}^{m}s_i=\mathcal{D}$, $|\mathcal{D}|$ is the number of instances in the data set $\mathcal{D}$.  
The set of data instances that is covered by considering all the rules in $R$ is also defined as $\mathcal{D}_R = \cup_{i=1}^{m}s_i$. Typically, $\mathcal{D}_R \subseteq \mathcal{D}$ because sometimes, there are instances just cannot be covered by any rule that qualifies the constraints.
\looseness=-1

The goal of the rule optimization procedure is to find the the minimal set of rules $R'\subset R$, that covers $\mathcal{D}_R$. This can be expressed as the classical \textit{set cover} problem\footnote{\url{wikipedia.org/wiki/Set_cover_problem}} which is proved to be NP-hard. To find an efficient approximate solution, we adapt the classic greedy set cover algorithm as follows: the algorithm adds one rule at a time to a target rule set we call $RS$ until every instance in $\mathcal{D}_R$ can be covered by at least one rule. Firstly, the algorithm adds the rule that covers the most instances. Then at each following step, the algorithm updates $uncover(i)$(the number of instances uncovered by $RS$ but by the $i$th rule) and chooses the rule that covers the most instances that remain uncovered to add to $RS$. The iteration continues until the whole $\mathcal{D}_R$ is covered or a stopping condition is reached (see below). The resulting $RS$ at the end of this process is the minimal set of rules that describes the model behavior on the training set we want to present to the user (the pseudo-code is included in the supplementary material). However, one potential problem with this procedure is that the algorithm may reach a stage where many rules that cover a very small number of instances are added to the rule set. To avoid this problem, we add a stopping criterion so that when $uncover(i)$ for unselected rules is smaller than $0.5\%$ of $|\mathcal{D}_R|$ the rules are discarded.  
\looseness=-1


\subsection{Hierarchical Surrogate Rule Set Construction}
\label{sec:hierarchy_ext}
Although the rules are extracted from a tree-based model, they are usually presented as a plain text list. To demonstrate the hierarchical structure, we need to extract the hierarchy from the rule set.

After extracting the rules from the trees we obtained from the random forest the next step is to ``stitch'' them back together in \textit{one} single new hierarchical structure. To this purpose, we construct an n-ary tree that is referred to as the \textit{Hierarchical Surrogate Rules (HSR)} which includes all the rules in the minimal rule set we get from the previous phase. The tree is built so that each leaf represents one of the original rules and each intermediate node represents a subset of the conditions that constitutes the rule. As will be discussed later in Section~\ref{section:vis}, the order of the conditions is meaningful. Therefore, the tree is built so that each level of the tree represents the position of the condition in the rule. The first layer of the tree contains all the conditions that appear in first position in the rules; the second layer adds the conditions that appear in second position in the rules and so on; adding conditions until all rules are represented in the tree. As an example, if the rule set includes the following four rules (with antecedent conditions $c_a, c_b, c_c, c_d$): $c_a \land c_b$, $c_a \land c_c$, $c_a \land c_d$, $c_c \land c_d$, the first layer will include two nodes representing respectively the conditions $c_a$ and $c_c$ and the second layer will include four nodes: $c_a \land c_b$, $c_a \land c_c$, $c_a \land c_d$, with $c_a$ as a common ancestor, and the condition $c_c \land c_d$ with $c_c$ as an ancestor. \looseness=-1


To build the tree from the rules, we first initialize the tree with an empty root node which will be the entry point for all the rules. Then we pick each rule one by one and add the conditions to the tree following the order of the conditions found in the rules. Because the same subset of conditions may be already present in the tree, we first find the longest common decision path and then add the new conditions. We repeat this procedure until we include all the rules in the rule set. The pseudo-code for constructing a hierarchical rule set is included in the supplementary material. 
In Section~\ref{section:vis}, we will discuss how to visualize the hierarchical rule set.

        
    
        
        

\section{Rule Set Evaluation}
\label{sec:algo_experiment}
In order to evaluate the algorithmic procedure, we presented above we describe two main experiments. The first, explores the behavior of the algorithm according to the parameters end-users can set at the level of individual rules: minimum \textit{rule fidelity}, minimum \textit{rule coverage}, maximum \textit{rule length}, and \textit{num\_bin}. More precisely, we investigate how these parameters influence rule set parsimony and completeness. Since the application has been designed to work in an interactive environment we also investigate how these parameters affect computation time. While the requirements for parameter settings do not need to be as stringent as to allow interactive updates when the parameters change, it is necessary to require the time to compute the results to be within a reasonable amount of time.
In the second experiment, we evaluate hierarchical surrogate rules (HSR), comparing it with surrogate decisions trees (SDT), a common solution used when building surrogate models.

All the tests described in this section are based on ten publicly available data sets from the UCI Machine Learning Repository~\cite{asuncion2007uci} and PMLB (Penn Machine Learning Benchmarks)~\cite{romano2021pmlb}. The data we use for the tests come from different domains and have been selected ensuring a wide diversity in terms of data size, dimensionality, and number of classes. Table~\ref{table:test_data} provides a summary of the main characteristics of the data sets we used. For all the experiments, we first used $80\%$ of the data set to train a classification model (using multi-layer perceptron or SVM) and reported test accuracy on the remaining $20\%$ of the data. Then, we fed the original training input (the $80\%$ portion) and the corresponding model output from the classification model to the algorithm with different parameter combinations we tested. More details can be found in the GitHub repository\footnote{\url{https://github.com/nyuvis/sure_evaluation}}, where the code for model training and experiments is included, as well as all the materials for the human-centered evaluation described in Section~\ref{sec:eval}.

\subsection{Sensitivity Analysis and Time Performance}
\begin{figure}
 \centering 
 \includegraphics[width=\columnwidth]{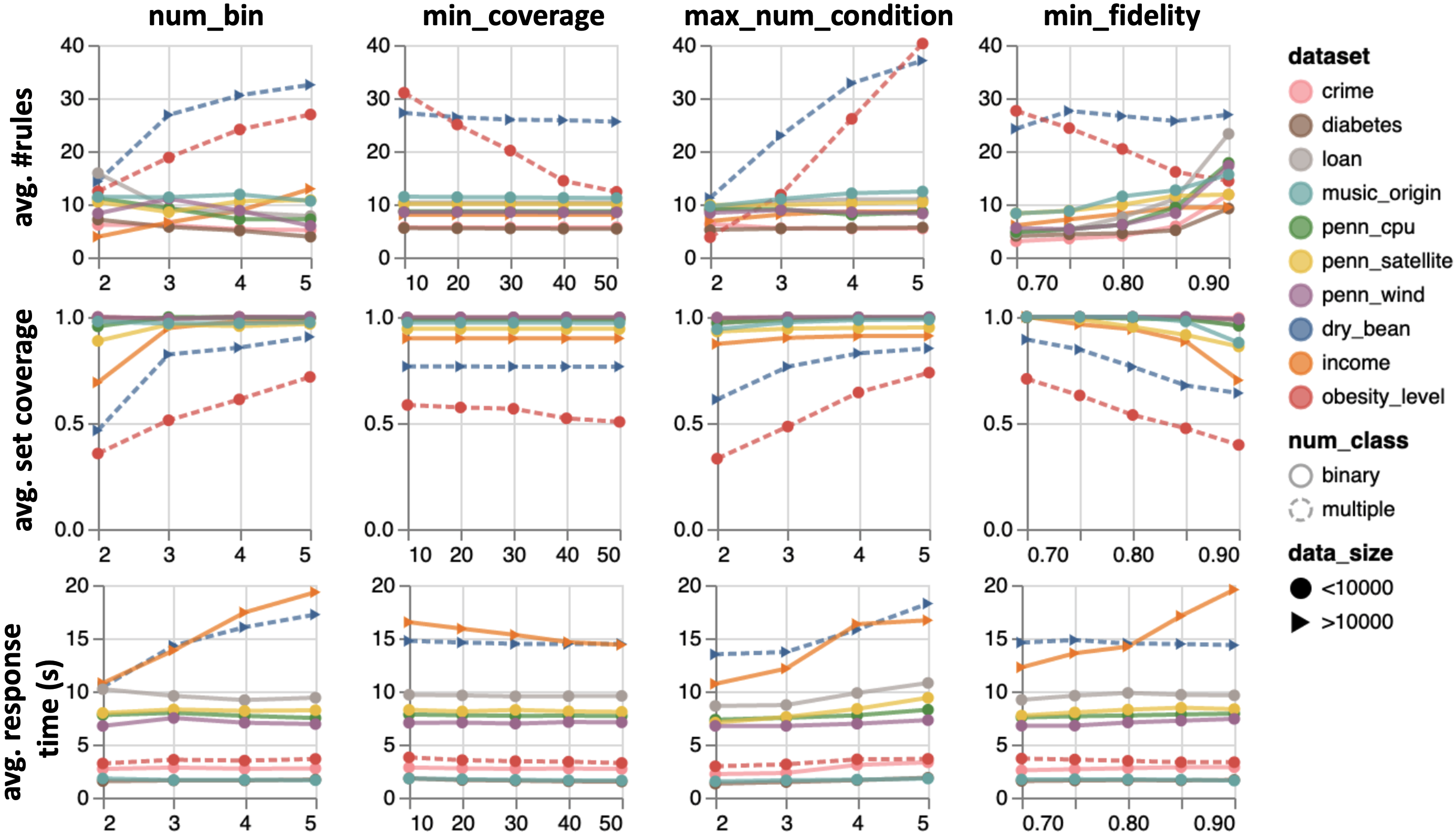}
 \caption{The experiments of evaluating how well the objectives can be reached and how effective the algorithm is.}
 \label{fig:experiment}
\end{figure}

The goal of the algorithmic procedure is to find a set of rules that maximizes completeness and parsimony. These two objectives are however influenced by how the end-user sets the rule generation parameters. For this reason, in this subsection we focus on analyzing how completeness (measured as \textit{set coverage}) and parsimony (measured as \textit{number of rules}) are influenced by these parameters. Also, given the relevance that interactivity has in a visual analytics setting, we include the analysis of how the parameters influence \textit{response time}. For all these tests we use the data sets outlined in Table~\ref{table:test_data} and show the results in Figure~\ref{fig:experiment}.

For the constrains, we explore the following value ranges. For \texttt{num\_bin} (feature granularity), values between 2-5 are set. For \texttt{min\_coverage} (rule coverage), values between 10-50 instances with a step size of 10 are used. For \texttt{max\_num\_condition} (rule length), values between 2-5 are used. For \texttt{min\_fidelity} (rule fidelity), values between $70\%$ and $90\%$ with a step size of $5\%$ are used.

To investigate the effect of these parameters on the generated rule sets, we calculate for each parameter value the average value of \textit{number of rules}, \textit{set coverage}, and \textit{response time} across the value ranges of all possible combinations of the other parameters. For example, when we calculate the average response time of the algorithm when \texttt{num\_bin} is 2, we run the algorithm by setting \texttt{num\_bin}$=2$ and then calculate the response time for all the possible combinations of \texttt{min\_coverage}, \texttt{max\_num\_conditions} and \texttt{min\_fidelity} within the range values we just described ($5\times4\times5=100$ combinations of the other parameters).



\begin{table}
    \centering
    \begin{tabular}{r r r r r r}
        \hline
        Data  & \#instance & \#feat & \#class & model & accuracy\\
        \hline
        \textcolor{text1}{crime}  & 1595 & 100 & 2 & MLP & $83.5\%$\\
        \textcolor{text2}{diabetes} & 1000 & 7 & 2& MLP & $75.5\%$\\
        \textcolor{text3}{loan} & 6317 & 21 & 2& SVM& $73.6\%$\\
        \textcolor{text4}{music origin} & 874 & \underline{117} & 2 & MLP & $80.7\%$\\
        \textcolor{text5}{penn cpu} & 6553 & 20 & 2 & MLP & $81.9\%$\\
        \textcolor{text6}{penn satellite} & 5148 & 35 & 2& MLP & $85.6\%$\\
        \textcolor{text7}{penn wind} & 5259 & 13 & 2& MLP & $83.2\%$ \\
       
        \textcolor{text8}{dry bean} & 10888 & 15 & \underline{7} & SVM & $93.1\%$ \\
        \textcolor{text9}{income}  & \underline{12545} & 5 & 2 & SVM & $76.5\%$\\
        \textcolor{text10}{obesity level} & 1688 & 15 & \underline{7} & MLP & $83.0\%$\\
        \hline
    \end{tabular}
    \caption{Details of data sets and the black-box models used for experiments. The number of instances (\textit{\#instance}) is only $80\%$ of the original set which is used for training the black-box model and a surrogate model.}
    \label{table:test_data}
\end{table}

\begin{figure*}
 \centering 
 \includegraphics[width=\textwidth]{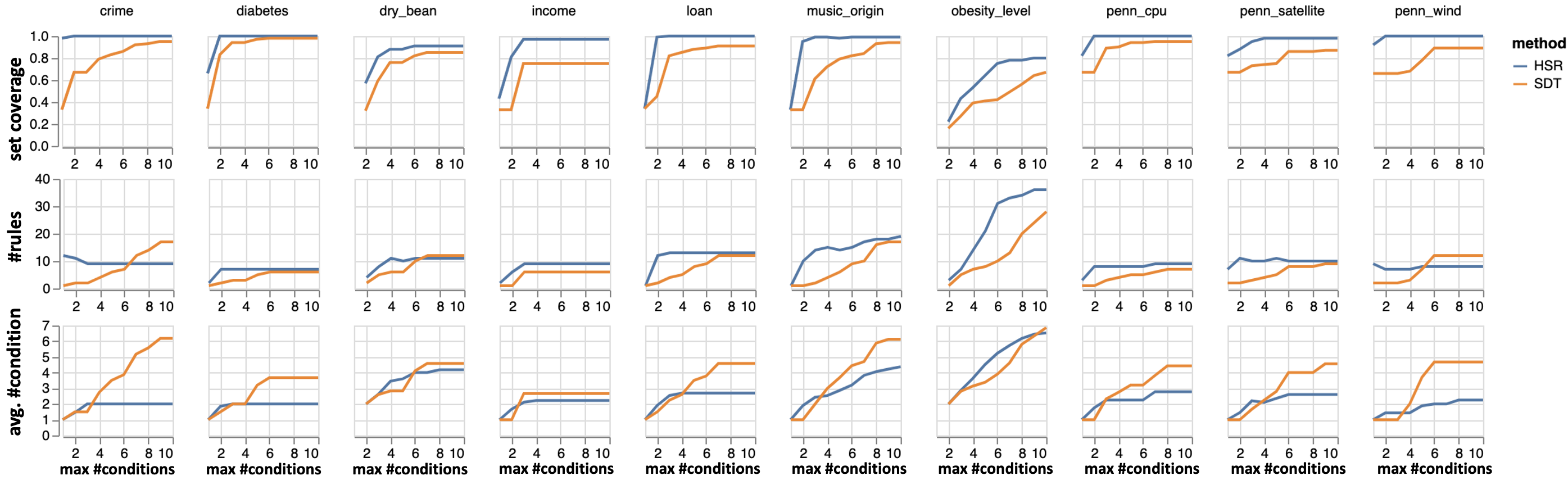}
 \caption{The comparison of rules extracted from one decision tree (SDT) and our approach (HSR). Here \textit{dry bean} and \textit{obesity level} are multiple class data sets. Both SDT and HSR cannot generate surrogate rules with only one condition to well approximate the model behavior for the multiple class data.}
 \label{fig:comparison}
\end{figure*}

The experiment results are shown in Figure~\ref{fig:experiment}. From left to right, the influence of different parameters is shown, including \texttt{num\_bin}, \texttt{min\_coverage}, \texttt{max\_num\_condition}, and \texttt{min\_fidelity}. From top to the bottom, the average values of i) size of minimal rule set, ii) data coverage by the rule set, and iii) response time (seconds) for the whole rule generation procedure are shown. The results are highlighted for the data sets of multiple classes using dashed lines, and large data sets with more than $10000$ instances using triangle marks. In the following part, the results are analyzed by how the parameters can influence the predefined objectives and the time performance.\looseness=-1

\textbf{Number of bins.} 
The increase of \texttt{num\_bin} does not have a strong effect on the number of rules overall. Only when the data sets are more complex, like more labels (e.g., \textcolor{text8}{dry bean} and \textcolor{text10}{obesity level}) or larger data sizes (e.g., \textcolor{text9}{income}), the increase of \texttt{num\_bin} leads to an increase in the number of rules. In some cases (e.g., \textcolor{text2}{diabetes} and \textcolor{text3}{loan}), however, more bins reduce the number of generated rules. As for the set coverage, the rule sets can cover almost $100\%$ of the data for most data sets except for the two multi-class ones. We also observe that the increase of \texttt{num\_bin} results in more coverage for more complex data sets, because the algorithm can capture more details when the \textit{feature granularity} is higher. In the column of \texttt{num\_bin}, we can clearly see a trade-off between complexity and completeness. Specifically, higher \textit{feature granularity} helps achieve higher coverage (more complete), although more rules are required to achieve that level of coverage (more complex). It is noted that even in the worst case among the benchmark data sets, the algorithm can always reach a reasonable coverage by around 30 rules on average, except for the \textcolor{text10}{obesity level} data set which requires even more bins and more rules to reach a higher data coverage. As for response time, the binning does not influence it too much: as we can see from the figure, most curves in the chart are flat. Only for the larger data sets (i.e., \textcolor{text8}{dry bean} and \textcolor{text9}{income}), the more bins, the more time is needed; which is still in a reasonable range given that it takes less than 20 seconds on average to compute in the worst-case scenario. 
\looseness=-1

\textbf{Minimum coverage.} In general, the value of \texttt{min\_coverage} does not have a strong impact on the metrics for most benchmark data sets. However, the results still show some diverging trends with multi-class data sets. With the increase of \texttt{min\_coverage}, each rule is required to cover more data instances, so the total number of rules decrease. There is no strong influence on the  number of rules and coverage for simpler data sets because the generated rules for these data sets can easily cover more than $50$ instances (the upper bound we tested). For \textcolor{text10}{obesity level} data, we also notice that although the number of rules decreases a lot with the increase of minimal rule coverage, the overall data coverage only drops slightly, which may be caused by the fact that some outlier instances cannot be described by a rule that covers enough instances. Lastly, as shown in the experiments, \texttt{min\_coverage} also does not influence the response time significantly.

\textbf{Maximum number of conditions.} For the binary classification data sets, the algorithm can use rules that contain no more than 2 conditions in each rule, while reaching a high data coverage. For multi-class data, however, with the increase of \texttt{max\_num\_condition}, the algorithm generates more rules with high enough fidelity in the rule pool, thus taking more time in both rule generation and optimization. In the worst-case scenario, the number of rules is around 40, which is still a reasonable number in terms of rules to be presented to the users. 
\looseness=-1

\textbf{Minimum fidelity.} Both \textit{number of rules} and \textit{set coverage} are more sensitive to the parameter of \texttt{min\_fidelity}. With the increase of \texttt{min\_fidelity}, the rules tend to be longer in order to reach higher fidelity so that each of them covers fewer instances. As a result of this, we can see that the algorithm generates more rules to describe as much of the data space as possible for the binary class data sets. However, the increase of \texttt{min\_fidelity} can also result in fewer rules that reach high enough fidelity, thus leading to relevant drops in \textit{set coverage}. In other words, when the constraint in fidelity is too high, reaching high coverage becomes more difficult. All the tested data sets have a major decline when going towards $90\%$ fidelity. For more complicated data sets, the shortage of high-fidelity (i.e., larger than the \texttt{min\_fidelity} threshold) rules also results in fewer rules in the rule set, which has a strong impact on the number of rules for multi-class data sets. Last, the effect of \texttt{min\_fidelity} on response time is minimal.
\looseness=-1

\subsection{Comparison with Decision Trees}

To generate surrogate decision rules that contain a hierarchical structure, the most straightforward solution is to use a single surrogate decision tree. In practice, this is how surrogate models are typically created in many real-world settings~\cite{augasta2012reverse,bastani2017interpreting,craven1996extracting,thiagarajan2016treeview,zilke2016deepred}. For this reason, in this subsection, we compare the rules generated from the HSR procedure to those extracted from a single surrogate decision tree (SDT). The goal of the comparison is to verify whether HSR can achieve more flexibility while retaining a useful hierarchical structure with reasonable complexity. More precisely, we want to see how \textit{parsimony} (\textit{number of rules}) and \textit{completeness} (\textit{set coverage}) compare when the two methods are constrained with the same parameters in terms of rule length, rule coverage and and rule fidelity.

To investigate this idea, we designed an experiment to compare the performance of the two solutions. More precisely, the experiment was designed to investigate these propositions: \looseness=-1

\begin{enumerate}
    \item \textit{SDT covers a smaller proportion of the data than HSR when they are both constrained to rules containing a small number of conditions (shorter tree depth).} Since HSR is able to use more combinations of conditions in the first layers of the tree structure, it is easier for HSR to cover more of the data with simpler rules.
    \item \textit{HSR generates a higher number of rules in order to simulate the original model effectively.} The higher flexibility in terms of combinations of conditions also means that HSR generates more rules than SDT in order to achieve higher coverage.
    \item \textit{SDT generates more complex rules (in terms of number of conditions) than HSR.} Since SDT is constrained to being a binary tree, the rules have to share many nodes/conditions from the early layers of the tree. Because of that SDT needs to generate longer and more complex rules.
\end{enumerate}

\begin{figure*}
 \centering 
 \includegraphics[width=\textwidth]{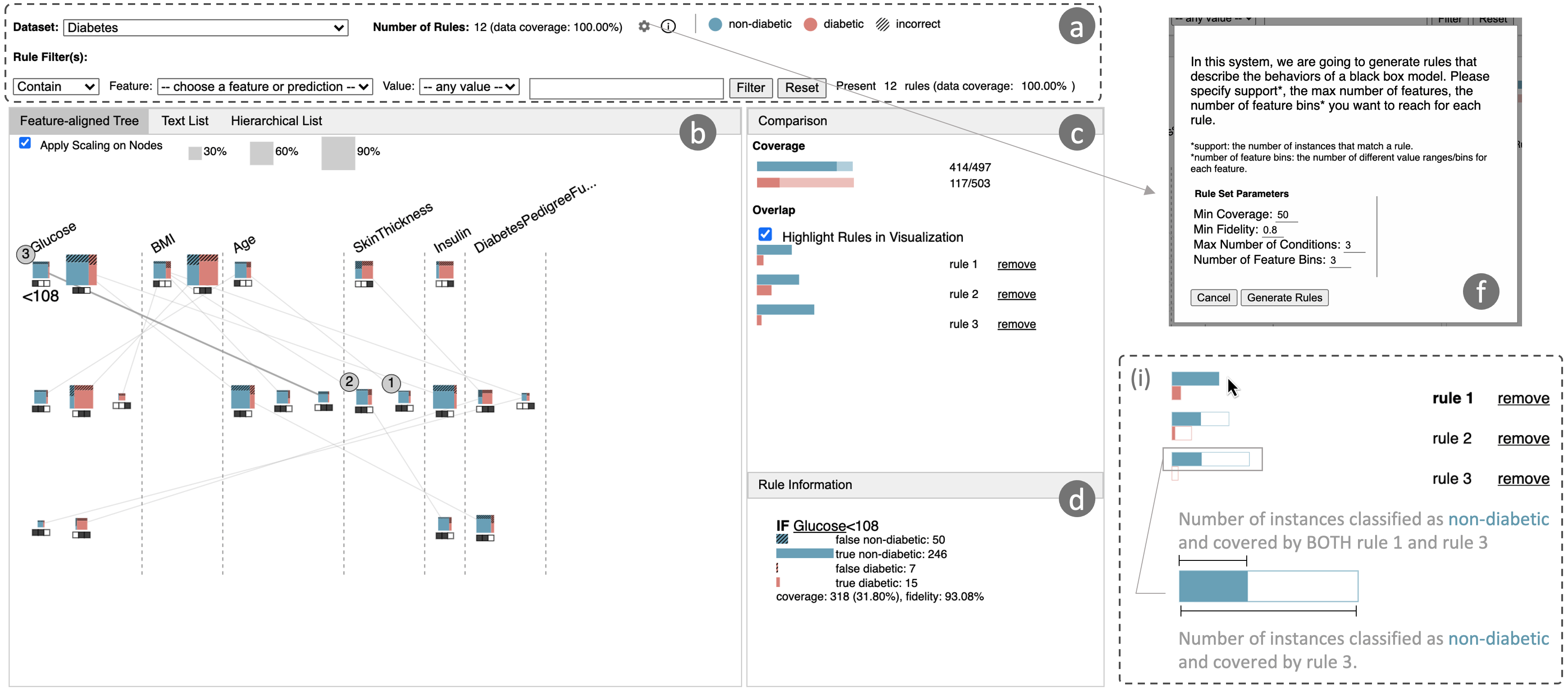}
 \caption{User Interface of SuRE contains five views: (a) Control Panel, (b) Rule Logic Visualization, (c) Comparison View, (d) Rule Information View, (f) Parameter Setting. When clicking on a node, this node will be given a number such as \textcircled{1}, \textcircled{2} and \textcircled{3}; the subgroup described by this rule is shown in (c) for further comparison. (b) contains three different visual designs of the same rule set. (i) shows that when a rule (rule 3 in the view) is hovered, the bars for other rules will also be updated to show the information of how many instances are overlapped in subgroups described by each rule and the hovered one. \looseness=-1}
 \label{fig:ui}
\end{figure*}

The experiment is designed as follows: (a) we set the same value of minimum rule fidelity, minimum rule coverage and number of bins for both HSR and SDT (\texttt{min\_fidelity} = $85\%$, \texttt{min\_coverage} = $5$, \texttt{num\_bin} = 3); (b) we vary the rule length constraint systematically (in a parameter sweep fashion) allowing the rules to have a maximum number of conditions between $1$ and $10$; (c) we compute the corresponding quality metrics: \textit{set coverage}, \textit{number of rules}, \textit{average number of conditions} in a rule. Note that in order to apply the same constraints to SDT and HSR we use the same procedure presented in Algorithm~\ref{algo:tree_pruning}. Specifically, HSR prunes all the tree paths in a random forest, and SDT prunes the paths from the trained decision tree. 
\looseness=-1

The results of the experiment are presented in Figure~\ref{fig:comparison}. In the figure, each column represents one tested data set and each row represents a quality metric. Each plot shows how the metric changes at different levels of maximum rule length for the two methods. An blue line is used for HSR and an orange line is used for SDT.

\textbf{Analysis of set coverage.} The first row of plots depicts the \textit{set coverage} metric. With the large majority of data sets, HSR achieves full coverage with a very small value of \texttt{max\_num\_condition} (often with as few as 2 conditions except multiple class data). Conversely, SDT achieves full or high coverage only with a much higher \texttt{max\_num\_condition}. Even when HSR is not able to cover the whole data set with a small number of conditions its level of coverage is always bigger than SDT (that is the blue line is always above the orange line).

\textbf{Analysis of number of rules.} The second row of plots depicts the \textit{number of rules} metric. As we expected, HSR generates more rules than SDT. In a way this is the fundamental trade-off between HSR and SDT. While HSR can achieve higher coverage with a smaller value of rule length (fewer conditions in a rule) it also needs to generate a higher number of rules in order to simulate the original model effectively. It is important to make two observations on HSR's behavior. First, the total number of rules it needs to generate for the data sets we tested is in general rather small. For all data sets except the one on obesity levels HSR needs around 15 rules maximum to simulate the model. Second, it is important to keep in mind that HSR can achieve a high level of coverage with shorter rules. Since the number of rules increases as the value of \texttt{max\_num\_condition} increases, it is possible for HSR to keep the number of rules at a similar level as SDT. In other words, if we let SDT grow to the level of complexity needed to have sufficient coverage, SDT also necessarily needs to produce a number of rules that is comparable to the number of rules generated by HSR (the orange line and the blue line converge or even have a cross when \texttt{max\_num\_condition} is high).

\textbf{Analysis of number of conditions.} The third and last row of plots depicts the \textit{average number of conditions} metric. This metric enables us to better understand how complex the rules generated by HSR and SDT are. As can be seen in the plots, as the constraints of \texttt{max\_num\_condition} is relaxed SDT tends to generate more complex rules than HSR. That is, while HSR is able to keep a small number of conditions in the rules, SDT tends to generate more complex (longer) rules. 
\looseness=-1

Overall, the experiment presented above shows that HSR provides a good compromise. It adapts better than SDT to the constraints imposed by the user. In particular it permits users to achieve high coverage, with a reasonable number of rules and smaller rule length. Although SDT can generate a smaller number of rules with respect to HSR, it sacrifices set coverage and uses longer rules. 

\section{SuRE: Surrogate Rules Explorer}
\label{section:vis}
Once the surrogate rules for a given model have been generated, it is crucial to have effective visual representations that enable the user to explore and make sense of them. Similarly, it is important to provide interactive means to change the parameters the algorithm uses to generate the rules and facilitate the analysis of rules.

For this reason, we designed and implemented a visual analytics system, {\sure}, which stands for \textbf{Su}rrogate \textbf{R}ules \textbf{E}xplorer\footnote{\url{https://github.com/nyuvis/SuRE}}. Figure~\ref{fig:ui} provides an overview of {\sure}'s user interface, which consists of five views to assist users in the process from rule generation to rule exploration and model understanding. The \textit{control panel} (Fig.~\ref{fig:ui}(a, f)) allows the users to set parameters for rule generation and to filter the rules according to a number of criteria. The \textit{rule logic view} (Fig.~\ref{fig:ui}(b)) visualizes the content of a rule set. The \textit{comparison view}(Fig.~\ref{fig:ui}(c)), enables the comparison of selected rules in terms of how much they overlap and what proportion of the data they cover. The \textit{rule detail view}(Fig.~\ref{fig:ui}(d)), shows more detailed information of selected rules (through hovering or clicking). All the views are interactively connected to display the information relative to the selected nodes/rules. 

In this section, we first focus on the main visualization we use for the hierarchical structure of the rules. We call this structure \textbf{Feature-Aligned Tree} because it uses a layout that aligns the nodes according to which features they include (see details below). Then, we describe the other views and the role they play in the analysis of the rules.

\subsection{Feature-Aligned Tree Visualization}

The structure generated by HSR is an n-ary tree with maximum depth smaller than or equal to the maximum number of conditions set by the end-user. To visualize this structure in a meaningful way we designed a feature-aligned tree like the one shown in Figure~\ref{fig:ui} (c). The tree layout is organized in rows and columns. Each row corresponds to a specific tree depth, from top to bottom. Each column represents a feature of the original data set. The nodes are placed in rows according to the tree depth level they belong to, and they are placed in columns according to which feature is used by the condition that has been added last to generate the node. Within each column the nodes are also ordered horizontally according to the values used by the condition; ordering low to high values from left to right.

Each node is represented by a small colored icon designed as follows (see the legend in Figure~\ref{fig:three_vis}). Size is proportional to how many data instances are covered by the rule. The icon is split horizontally into as many segments as the number of labels in the data (two labels in the figure). Each segment is colored with a different hue and is proportional to how many instances in the rule have a specific label. Within each horizontal split there is an additional vertical split that segments the area to represent the number of instances that are predicted correctly (plain color) or incorrectly (shaded part) by the original model. Finally, the nodes are connected by lines that represent the sequence of conditions of each rule. For example, in Figure~\ref{fig:three_vis}(c), the node\circled{1} represents the set of conditions $Insulin<115 \land Glucose<136 \land Age<37.3$ \footnote{Keep in mind that even though we show numeric values as splits in the conditions of these rules, they are all discretized into the user-selected number of bins}. When the user hovers over a node, the path that goes from the node back to the root through the list of parents is highlighted and the textual representation of the selected rule is presented in the bottom right section of the interface. All the paths that lead to the children of the selected node are also highlighted.
\looseness=-1

The main characteristic of the feature-aligned tree is to have the nodes that use the same feature grouped in the same column. The feature-aligned tree allows quickly singling out all the conditions and rules that use a given feature. This is important because often one wants to see what the effect of a given feature is on the prediction. Since the features are all aligned it is easy to compare them and to relate the conditions they have with the label distribution they generate (depicted by the colored areas). The tree visualization also has several advantages that derive from being organized in layers that correspond to rule length. This organization permits users to see what is the impact of conditions as they become progressively more complex and to find a reasonable trade-off between complexity, fidelity and coverage. 
Also, following the paths from a given node down to its children it is possible to quickly grasp how the class distribution changes and how it is influenced by the features used by the child nodes.
\looseness=-1

Finally, the design of the colored icon also enables several relevant tasks. It allows the users to: identify nodes with a given class distribution (typically one skewed towards one label since it represents more purity and thus higher fidelity); identify nodes with more errors from the classifier (longer shaded areas); and relate the value of the condition with the class distribution (e.g., high \textit{Insulin} leading to more predictions of diabetes).

Since this visual representation is novel and unfamiliar to most users, it is important to verify that the visualization can be quickly learned and used appropriately. Therefore, in Section~\ref{sec:eval_fat} we present a usability study which aimed to verify to what extent novice users can learn how the visualization works and carry out a series of fundamental tasks. 
\looseness=-1

\subsection{Text-Based Visualizations}
\begin{figure}
 \centering 
 \includegraphics[width=\columnwidth]{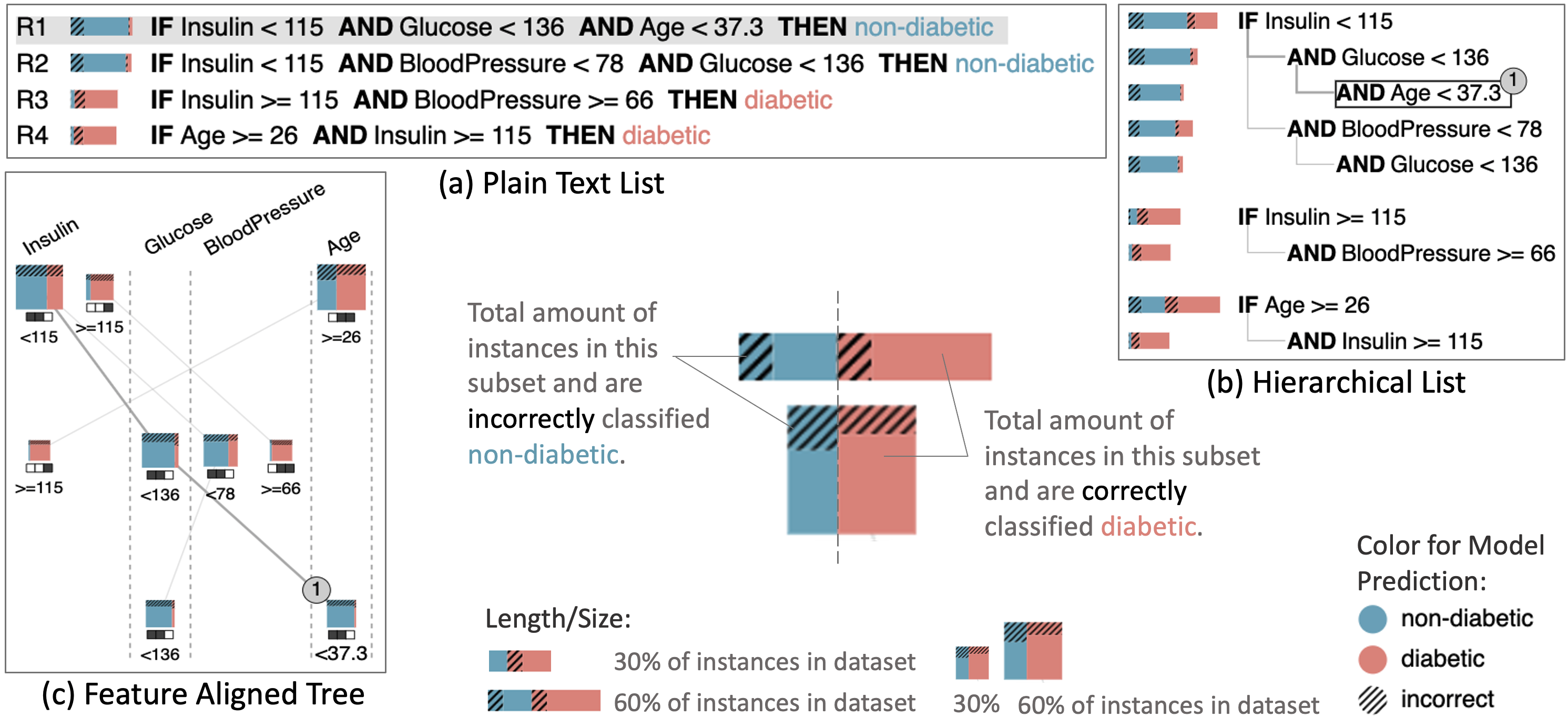}
 \caption{Rule logic visualization designs: (a) conventional plain text list, does not contain hierarchical structure; (b) hierarchical list; (c) feature-aligned tree design. }
 \label{fig:three_vis}
\end{figure}

In addition to the tree visualization we included in {\sure} two additional visualization based on text. The first showing the list of rules extracted from the algorithm, the \textit{Text List} view, and second showing the hierarchical structure of the rules through a hierarchical list, the \textit{Hierarchical List} view. Figure~\ref{fig:three_vis}(a,b) shows the two visualizations. The text list is the traditional textual list of logical statements found in many existing systems augmented with a series of small bars showing rule coverage (size of the bar), label distribution (size of the colored segments) and error distribution (size of the shaded segments). The hierarchical list has the same graphical design solutions of the plain text list with the additional feature of indenting the rules according to the hierarchical structure obtained from the algorithmic procedure.
\looseness=-1

We introduced these additional visualization for two reasons. First, to make sure our users can find the most familiar representations they expect from existing systems. Second, to enable a smoother transition from the most classic views to the more advanced tree-based visualization. In addition to that, having these separate views also allowed us to observe, in the study we present in Section~\ref{sec:eval}, real-users decide which one to use according to their analytical intent. As described in Section~\ref{sec:eval}, being able to observe how users analyze rules with different visualizations and to ask feedback from our study participants allowed us to derive interesting observations.

\subsection{Interactive Rule Generation and Exploration}
\label{sec:ui_interaction}
The user interface of {\sure} enables users to adjust a few constraints introduced in Section~\ref{section:pipeline}. This can be done by clicking the gear icon in the control panel which opens the pop-up window shown in Fig.~\ref{fig:ui}(f).

As the user explores the results, the user interface provides additional information pertaining to the selected rules/node in the auxiliary panels on the right-side of the interface Fig.~\ref{fig:ui}(c, d). The \textit{rule information} panel (bottom-right) provides detailed information about the selected rule, including a textual description of the rule and a breakdown of the distribution of the data covered by the rule in terms of the labels and the model errors. The \textit{comparison} panel (top-right) enables rule comparison when multiple rules are selected in the visualization. The panel shows the number of data points of each label covered by each rule (see the colored bars). When the user hovers over the bars of one rule, the bars of the other rules show the overlap between the selected rule and the others (see the explanation in Fig.~\ref{fig:ui}(i)). \looseness=-1

During the rule exploration, {\sure} also enables users to inspect a subset of rules by filtering (as shown in Fig.~\ref{fig:ui}(a)). Users can filter rules by \textit{feature existence}, \textit{feature value}, and \textit{model prediction}, which allows the inspection of rules that are relevant to specific information a user is interested in.

\section{Evaluation}
\label{sec:eval}
{\sure} integrates interactive rule generation and exploration, as well as unique visual designs for the representation of rule logic. In this section, we describe two human-centered evaluations we carried out to validate the tree visualization we developed as well as the whole interactive system.

In the first study, we evaluate the readability and comprehensibility of the feature-aligned tree visualization through a usability study. We then report on the qualitative evaluation results of the observational study with seven data scientists and domain experts from a banking company.

For both evaluations, we use two data sets: a diabetes data set~\cite{smith1988using} (\textcolor{text2}{diabetes} data set in Table ~\ref{table:test_data} to predict whether a person is diabetic or not) in the training phase to help the participants familiarize themselves with the visualization and tasks, and a much more complex and realistic credit risk data set~\cite{fico} (\textcolor{text3}{loan} data set in Table ~\ref{table:test_data} to predict whether a loan applicant will default or not) for the actual study. 

\subsection{Evaluation of Feature-Aligned Trees}
\label{sec:eval_fat}
The goal of this study is to test whether users can accurately extract and search for information from the proposed feature-aligned tree visualization. To this purpose, we conducted a task-based usability study in which we assigned several prototypical tasks and recorded information about error rate and perceived effort. In the following subsections, we describe the study design and the results.

\subsubsection{Tasks}
\begin{figure}
 \centering 
 \includegraphics[width=\columnwidth]{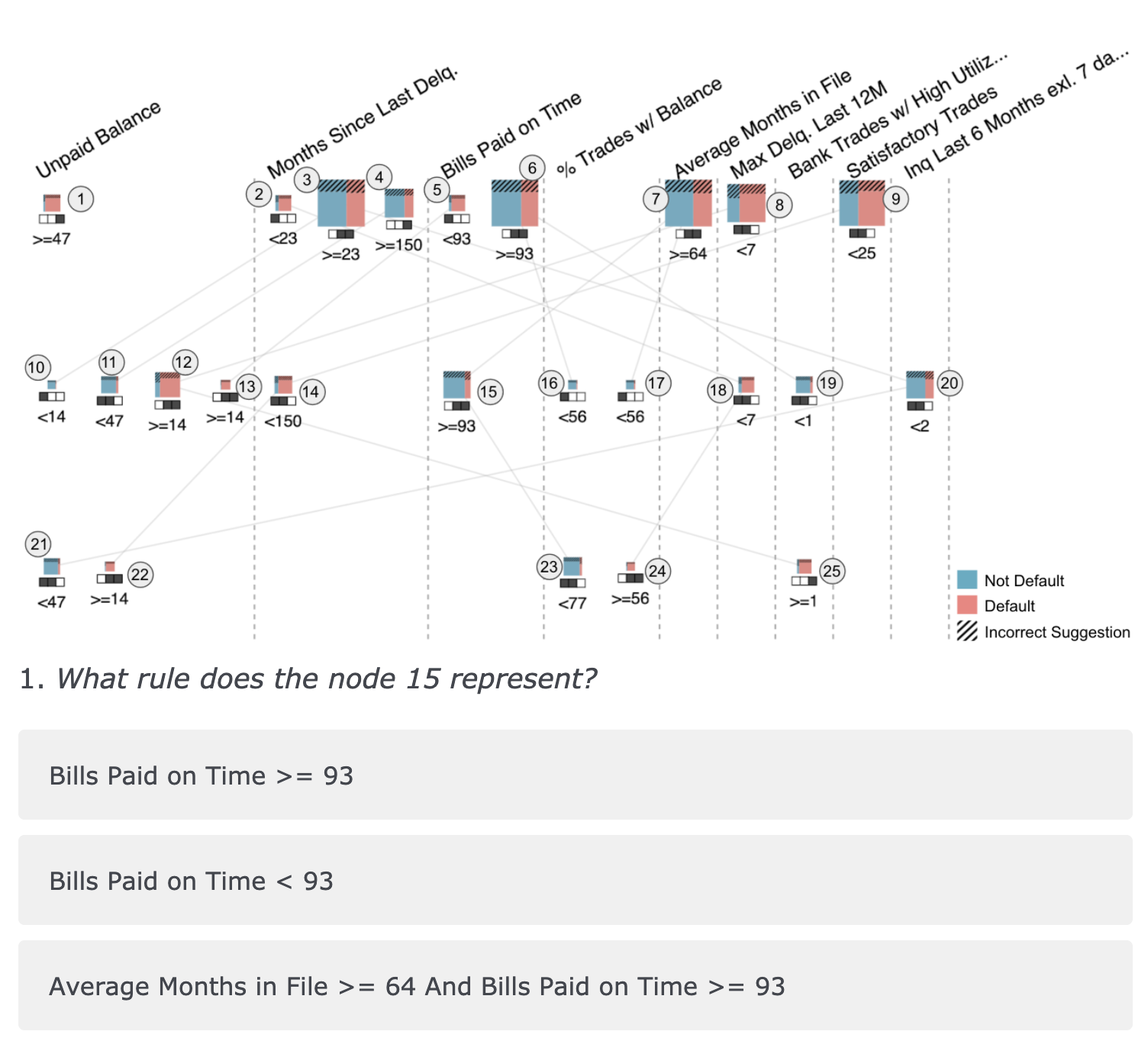}
 \caption{The screenshot of the question of task T1 in the evaluation of the feature-aligned tree.}
 \label{fig:usability_t1}
\end{figure}

In designing the tasks for the study, we identified two broad sets of tasks that simulate the way people engage with rules:

\begin{itemize}
    \item[\textbf{T1}] \textbf{Visual Interpretation.} This category of tasks focuses on testing whether the user can correctly interpret the information presented in the visualization. In this category of tasks, we included questions to test the interpretation of the rule conditions, rule-related statistics (e.g., coverage, error rate), and the outcome of rules. An example question to demonstrate this is shown in Fig.~\ref{fig:usability_t1}, where we assign numbers to the nodes in the feature-aligned tree and ask the participants to interpret the rule content of a node. \looseness=-1
    
    
    \item[\textbf{T2}] \textbf{Visual Search.} This category of tasks intends to simulate the situation where the user looks for specific elements of interest. For example, a data scientist may be particularly interested in the rules that have a high level of purity in a node, or rules that use a specific feature. To simulate these cases, we ask the participants to search for a node that matches a given description and to select it once they have found it. For example, one of the questions we asked is "Please select the node that covers the least number of applications." 
\end{itemize}

All the questions we asked in the actual test are listed in the Table ~\ref{tab:study1_questions}. 

\begin{table*}[]
    \centering
    \begin{tabular}{|l|l|}
         \hline \textbf{Task Type} & \textbf{Test Questions} \\
         \hline 
         \multirow{5}{*}{\makecell[l]{Visual \\Interpretation }} & *``What rule does the node 15 represent?'' \\
          & ``What rule does the node 14 represent?''\\
          & ``What is the most probable outcome/prediction of the model for applications that match the rule represented in node 18?''\\
          & ``Which of the following nodes covers the most people?''\\
          & ``Which suggestion has higher error rate with node 8?''\\\hline
          \multirow{7}{*}{Visual Search} & *``Please click a node that suggests \textit{Not Default}.''\\
          & ``Please click a node that contains a condition with \textit{Month Since Last Delq.}''\\
          & ``Please click a node that contains a condition with \textit{Bills Paid on Time} and suggests \textit{Not Default}.''\\
          & ``Please click a node that presents a condition of \textit{Months Since Last Delq.}$>=150$ AND \textit{Unpaid Balance}$<47$.''\\
          & ``Please click a node that you think it covers the least number of applications.''\\
          & ``Please click a node whose \textit{Not Default} suggestions have a higher error rate than \textit{Default} suggestions.''\\
          & ``Please click a node that represents \textit{Months Since Last Delq.}=Low, AND \textit{Max Delq. Last 12M}=Low OR Medium.''\\\hline
    \end{tabular}
    \caption{The test questions in the two types of tasks involve rule conditions, rule related statistics, rule outcome. Questions with a star (*) symbol are only for familiarization and not considered in the result.}
    \label{tab:study1_questions}
\end{table*}

\begin{table*}[]
    \centering
    \begin{tabular}{|l|l|l|l|l|l|l|l|l|l|}
    \hline
    \multicolumn{4}{|c|}{\textbf{T1. Visual Interpretation}}  & \multicolumn{6}{c|}{\textbf{T2. Visual Search}} \\\hline
      Q1 & Q2 & Q3 & Q4 & Q5 & Q6 & Q7 & Q8 & Q9 & Q10 \\\hline
      \green{24/24} & 23/24 & 23/24 & 23/24 & \green{24/24} & 23/24 & 23/24 & 22/24 & \green{24/24} & \green{24/24} \\ 
    \hline
    \end{tabular}
    \caption{Task performance. The first value in each cell is the number of participants that answered this question correctly, while the second value, 24, is the total number of participants. Most participants answered all questions correctly. \looseness=-1}
    \label{tab:usability_test}
\end{table*}

\subsubsection{Performance Measurements}
To evaluate the performance of both tasks, we use error rate as the main performance parameter. The focus here is to verify that users can extract information from the feature-aligned tree correctly. For T1 tasks, we simply check whether participant chooses the correct answer. As for T2 tasks, since there may be more than one node  that matches the given description, we count the answer as correct if all the clicked nodes match the description. 

We purposely designed this study to exclude the results of the first question of each task type to count for habituation effects (the results of these tasks are available in the GitHub repository mentioned above).

To evaluate \textit{perceived effort} we also asked the study participants to report their subjective sense of effort for each type of tasks. This score was measured at the end of each task type by participants answering the question of ``To what extent do you agree with the following statement: \textit{It was easy for me to answer the questions.}'' To analyze the responses, we report the distribution of the responses for both task types.

\subsubsection{Participants and Study Procedure}
A total of 24 volunteers (2 undergraduate students, 22 graduate students; 14 male, 10 female) were recruited to participate in the study. We first sent an email to all students currently enrolled in the Computer Science and Engineering department as well as the Center of Data Science from New York University. Students who indicated they were interested in this study were then asked to fill out a form. Then, we sent follow-up emails to those who had experience with machine learning (as was indicated by the results of their survey) and scheduled the studies with them.

We organized 4 online/remote meetings for the study. In each meeting, there were between 3 and 9 participants. At the beginning of each study, we first collected the consent forms and then moved on to describing the visualization and explaining how to read it. The tasks were then introduced, and we explained what each task meant and how to carry them out using the visualization. At each of these steps, we encouraged the participants to ask questions and provided further clarifications when needed.

In order to verify the participants' understanding of the tasks, we then asked the participants to answer a few questions using the visualization with the diabetes data set. After engaging with the training tasks, we provided further help and clarifications when needed. Finally, the participants were directed to a survey containing the tests that the participants had to perform individually. During this time, the participants were required to mute themselves and they were not allowed to communicate with other participants. The participants were only allowed to direct message us for any questions related to the test but not to get help in carrying out the task.
\looseness=-1

\begin{figure}
 \centering 
 \includegraphics[width=\columnwidth]{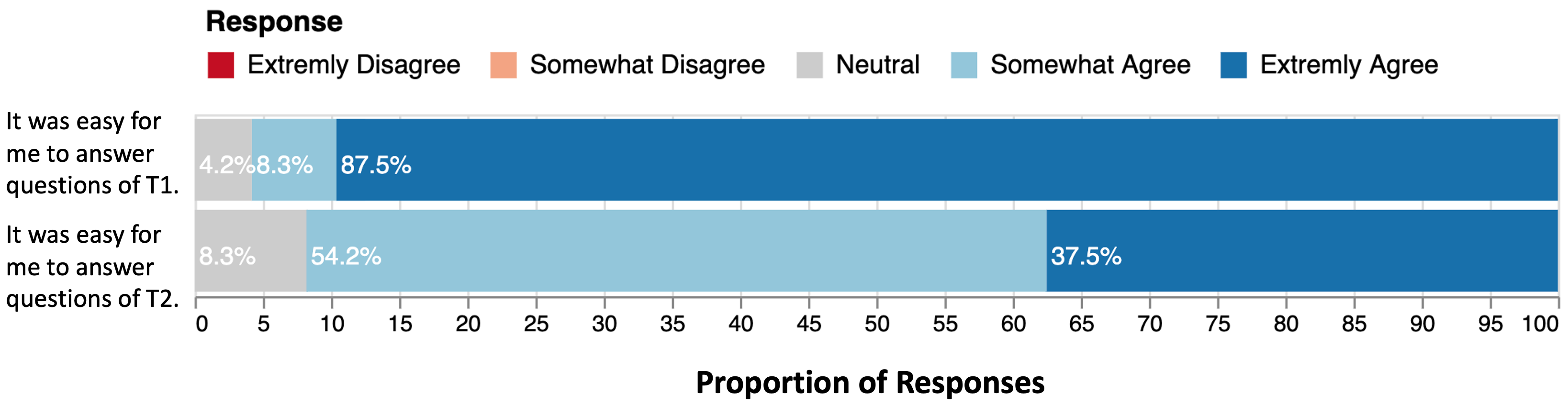}
 \caption{The proportion of responses to the question of ``To what extent do you agree with the following statement''.}
 \label{fig:usability_easy}
\end{figure}

\subsubsection{Results}
For both T1 tasks and T2 tasks, most participants were able to answer all the questions correctly. As shown in Table ~\ref{tab:usability_test}, most participants answered all questions correctly. For questions Q2, Q3, Q4, Q6, Q7 and Q8 (6 out of 10 questions) accuracy was only slightly lower than $100\%$.

As for perceived effort, most participants agreed that the two task types were easy for them ($95.8\%$ for T1 and $91.7\%$ for T2, see  Fig.~\ref{fig:usability_easy}). For T1 tasks, $87.5\%$ of participants extremely agreed that the tasks were easy to answer. For T2 tasks only $37.5\%$ of participants extremely agreed that the tasks were easy to answer. This makes sense because these tasks not only required participants to read the information from the visualization but also to compare and search desired visual elements in the view. We also noticed that no participant thought the questions for both task type T1 and task type T2 were not easy given this new visualization.
\looseness=-1

\subsection{Evaluation of {\sure}} 
\label{sec:eval_sure}
To evaluate the utility of {\sure}, we designed an observational study with a group of data experts. The goal was to observe experts engage, for an extended period of time, in data analysis tasks using {\sure} in order to understand how the data experts think about model understanding and evaluation problems, and to what extent {\sure} supports their way of thinking about these problems. In this sense the outcome of the evaluation is not exclusively about verifying that {\sure} supports the experts' tasks, but also to shed a light on what tasks experts need to perform and how they could be supported further with future solutions. In other words, rather than assigning a set of predefined benchmark tasks, we simulated a high level model analysis scenario (see details below) and observed what specific analytical tasks would emerge from the analysis, and to what extent these tasks were supported by {\sure}.


\subsubsection{Participants}
We invited 7 practitioners (3 female, 4 male) from a banking company to participate in this study. As shown in Table~\ref{table:interviewee}, our participants have a varying degree of expertise in industry, including participants who just started their career (P1, P2, P3) and those who worked with data for more than 10 years (P4, P5). All participants have a strong background in data science and they cover different job titles, including: \textit{data scientist}, \textit{software engineer}, \textit{research scientist}.  \looseness=-1
 
\begin{table}
    \centering
    \begin{tabular}{|c|c|c|c|}
    \hline
         \textbf{Interviewee} & \textbf{Degree} & \textbf{Experience Level} & \textbf{Job Title} \\\hline
         P1 &  Master & entry-level & Data Scientist \\\hline
         P2 & Bachelor & entry-level & Software Engineer \\\hline
         P3 & PhD & entry-level & Research Scientist \\\hline
         P4 & PhD & senior & Research Scientist \\\hline
         P5 & PhD & senior & Research Scientist \\\hline
         P6 & PhD & junior & Data Scientist \\\hline
         P7 & Master & junior & Data Scientist \\\hline
    \end{tabular}
    \caption{We conducted an observational study with domain experts from a banking company. Among all the participants, only P2 is a software engineer and was working on the development of an machien learning platform. The rest of the interviewees are data scientists and researchers who work on model development and data analysis.}
    \label{table:interviewee}
\end{table}

\begin{table*}[]
    \centering
    \scalebox{0.96}{
    \begin{tabular}{|l|l|l|l|l|l|}
        \hline
        \makecell[l]{\textbf{Analytical} \\ \textbf{Focus}} & \textbf{Task Category } & \textbf{Count} & \textbf{Description} & \textbf{Example Question/Task} & \makecell[l]{\textbf{{\sure}} \\ \textbf{Support}}  \\\hline
        
         \multirow{2}{*}{\makecell[l]{Search/\\Explore}}  & \makecell[l]{Rule Search\\ / Exploration} & 11 & \makecell[l]{Search for rules with given properties. \\and see if something stands out} & ``What are the nodes with high fidelity?'' & Full \\\cline{2-6}
         
        & Feature Relevance & 6 & \makecell[l]{Understand feature relevance /Look \\ for features with a given relevance} & ``What are the most important features?'' & Full \\\hline
         
          \multirow{3}{*}{Explain} & \makecell[l]{Feature-Outcome \\ Logic} & 11 & \makecell[l]{Understand the role one or more features\\  play with regard to the outcome.} & \makecell[l]{``What is the relationship between one\\  feature and the outcome in general?''} & Full\\\cline{2-6}

          & \makecell[l]{Outcome-Feature \\Logic} & 10 &  \makecell[l]{Understand what leads to a given outcome.} & \makecell[l]{``What features are used \\in common for class 1?''} & Full \\\cline{2-6}

         & \makecell[l]{Feature-Error\\ Logic} & 3 & Understand what leads to errors. & \makecell[l]{``Whether the usage of a feature \\ increase or decrease errors?''} & Partial \\\hline

         \multirow{2}{*}{Reconstruct} & \makecell[l]{Threshold\\Adjustment}& 6 & Explore different feature value thresholds. & \makecell[l]{``Whether the feature value change \\can lead to different prediction?''} & None \\\cline{2-6}
         
         & Model Search& 6& \makecell[l]{Create a new set of rules to satisfy\\ certain desired properties.} & ``Can I check simpler rules?'' & Full \\\hline
    \end{tabular}
    }
    \caption{We categorized the tasks the domain experts performed when using rules to understand model behaviors.}
    \label{tab:task_characterization}
\end{table*}

\subsubsection{Procedure}
Each observational study lasted between $90$-$120$ minutes and was conducted through an online meeting using a videoconferencing platform.
We began each session with an introduction, during which we clarified the goals of the study, showcased {\sure} and explained how to read its three different rule visualizations ((a), (b) and (c) in Fig.~\ref{fig:three_vis}). We provided a brief interactive tutorial regarding the usage of {\sure}. Similar to what we did in the aforementioned usability study on feature-aligned tree, after this initial tutorial we asked the participants to fill out a survey to verify whether they understood how to read the rule visualizations properly. Then, the participants were asked to explore {\sure} freely using the diabetes data set. To make sure the participants were able to make full use of the interactive functionalities, we asked them to answer two open-ended questions during the exploration: (1) \textit{``What is the most probable outcome when a person has a low/medium/high value of BMI?''} and (2) \textit{``Given a person is predicted as having diabetes, what characteristics may this person have?''} \looseness=-1

After this initial training phase, we moved on to the actual study. For the study we tried to simulate a real-world scenario. The participants were told that their goal was to explore a pre-trained binary classifier based on loan applications data~\cite{fico} and to produce a summary report describing how the model behaved. In this step, we instructed the participants to follow a ``think-aloud'' protocol~\cite{fonteyn1993description}, that is, participants were required to reason out loud and to explicitly mention what questions they were trying to answer. \looseness=-1

During the analysis, the participants shared their screen with us so that we could observe all their interactions. The shared screen, their camera feed and their voice were recorded to enable further analysis after the session. Two of the authors also took notes while observing the participants' behavior and thinking. Specifically, the following notes were taken: (1) the questions the participants wanted to answer for model understanding, (2) to what extent it was easy to answer the questions with {\sure}; and (3) how the participants interacted with {\sure} to answer the questions.

The analysis session was then followed by a debriefing session in which we reviewed the results with the participant individually. We conducted a semi-structured interview which incorporated several questions about visual design, overall usefulness, general pros and cons of {\sure}, as well as how they imagined such workflow could be applied to their own data or in their own work.
\looseness=-1

Once all the sessions were finished, we gathered all the notes and recorded material to analyze the results. The analysis focused on the questions collected from all the participants during the study. Based on the notes, two authors \textit{independently} coded a total of 63 questions. The intent was to group the tasks in several categories to capture major task types. The two coders then met and compared categories to find commonalities and disparities. The coders first identified categories with similar or same meaning and resolved conflicts at the level of categories. After agreeing on a group of categories, the coders analyzed cases of questions that were coded with different categories and found ways to resolve the conflict; either by agreeing on a category or setting the question aside as a special case. At the end of this iterative approach, we reached a consensus on 7 categories and on the assignment of 53 questions to these categories. The 10 questions that were not included were either out of scope (5 out of 10) or covered special one-off cases which did not make sense to report as a whole separate category (intermediary results can be found in the supplementary material). In Table~\ref{tab:task_characterization}, we report the seven major task categories derived from the collected data. \looseness=-1




\subsubsection{Task Characterization and Support}

The main outcome of the coding activity is a characterization of the tasks we observed and the level of support {\sure} provides to these tasks.



As shown in Table ~\ref{tab:task_characterization}, the tasks collected during the observational study can be categorized into seven categories, which were then grouped into three analytical focuses. 

The first analytical focus \textit{search / explore} pertains to tasks that revolve around searching for elements of interest and/or making sense of the elements observed in the visualization. This analytical focus includes two main task categories.
The first category \textit{Rule Search / Exploration} involves the search and exploration of a rule with desired properties. For example, P4 looked for nodes (rules) with a high error rate in the visualization to understand ``where the model does not work well''. The second task category \textit{Feature Relevance} involves tasks about exploring features according to their relevance. For example, P5 looked for features with more and larger nodes in the column in the feature-aligned tree view to understand what features are more ``important''. {\sure} provided good support to the two task categories. All the participants were able to find the desired information from the visualization either directly or with the help of filtering to narrow down the search space.

Another analytical focus is to \textit{explain} the presented rule logic to learn the specific model behavior, which includes three major tasks. The first is \textit{Outcome-Feature Logic}. This category describes all the tasks that focus on explaining what feature values led the model to make a certain type of prediction (rule outcome).
For example, P1 asked ``which features are in common for default/not default?'' while reasoning about the rules filtered by class. The second category, \textit{Feature-Outcome Logic} focuses on the reverse type of task, that is, how the values of a feature of interest influence the rule outcome, which represents the model predictions. For example, P5 worked with financial data for a few years and was interested in the feature called ``avg. months in file'' based on their prior knowledge, so P5 checked how this feature influenced the model prediction. Lastly, some domain experts also expressed interest regarding where and how the model made mistakes, so they performed tasks of the type referred to as \textit{Feature-Error Logic}. This category contains the tasks of explaining the relationship between given feature values and the error rates. For example, P5 tried to reason about the relationship between a presented feature range and the error rate. P5 also followed a decision path in the feature-aligned tree to understand ``whether the usage of a feature increase or decrease the errors.'' In general, {\sure} provided full support to \textit{outcome-feature logic} and \textit{feature-outcome logic} tasks, but only supported the \textit{feature-error logic} tasks partially because the rules were generated to approximate model prediction behavior but not erroneous behavior. \looseness=-1

The third analytical focus is \textit{Reconstruct}. In this focus the goal is to reconstruct the rules to understand the model behavior using different descriptions. 
The \textit{Threshold Adjustment} tasks are related to changing the threshold of feature values used in the rule to understand the influence of feature value changes and/or validate the presented threshold value. The questions in the task category usually starts with ``what if''. For example, P1 wanted to know what if the threshold value changed in specific rules. P3 also wanted to check ``whether the feature value change (in a condition) can lead to different predictions''. At present, {\sure} does not provide support to these tasks. However, we found \textit{threshold adjustment} was useful to validate the influence of specific features on outcome or errors. So we are highly interested to support this in the future. Another task category to \textit{reconstruct} rules is \textit{model search}, which is about creating an entirely new rule set with desired properties (e.g., \textit{min\_num\_conditions}, \textit{min\_fidelity}, etc.) to approximate the model behavior. For example, in the beginning of P2's session, P2 generated a simpler rule set with few rules by setting lower \textit{min\_fidelity} to gain a general idea of model behavior. During the observational study, we found that creating a new rule set was well supported by {\sure}, with users not having any specific troubles and setting the desired parameters and generating the associated rule sets.

An interesting aspect of the tasks defined above is that they are not necessarily independent. Often, we noticed that our participants carried out a task that is actually a combination of the basic tasks we described. Two combinations seemed to take place with some frequency.


\textit{Search/Explore $\rightleftharpoons$ Explain:} In some cases, domain experts had some features of interest based on their prior knowledge or the information they have gained from the system. Therefore, they \textit{search / explore} the rules involving the feature they have in mind, and then \textit{interpret} the corresponding logic information. Reversely, after interpreting some presented rule logic, the domain experts can continue the search and exploration with other rules, features, class, etc. They may also search the same element of interest with other properties. \looseness=-1

\textit{Search/Explore/Explain $\rightleftharpoons$ Reconstruct:} After gaining some understanding of the model behavior, the domain experts either wanted to change some rule values manually to further test their hypothesis or generate a new rule set with some constraints changed, for example, longer but higher-fidelity rules. In turn, once the reconstructed rules were presented, the domain experts repeat the tasks in the other task categories to continue the analysis of model behavior.

\subsubsection{Observed Behavioral Patterns}

Through our observations and interactions with the participants, we identified a few relevant behavioral patterns. Such patterns are not only interesting in the context of the development of {\sure}, but also more in general in identifying behaviors that have interesting implications for the development of visual analytics systems for model exploration.
\looseness=-1

\textbf{Complementary role of rule visualizations.} During the study, we found that participants tended to switch between the three visualizations of rules we provided in {\sure}. This came to a surprise to us. surprising. Our initial expectation was that the participants would find a preferred view and keep working with it. But the participants kept switching between views and clearly found them useful at different times for different tasks. After the session, we asked the participants why they used all the available views and what they thought their relative strengths were. It turns out the three views do seem to provide complementary strengths. According to participants, the \textit{Text List} made it easy to just focus on the actual set of rules extracted by the algorithm. Some used the list as a starting point, then focused on specific features and analyzed their roles in the other views. P2 said ``\textit{The text list is good at checking the actual rules.}'' P6 also commented that ``\textit{text list is good to compare the distribution [of the complete high-fidelity rules] and see how many rules are involved.}'' The list was also a good default to a representation that is familiar to most users. P3 mentioned there was still a learning curve for other visualizations. And P6 said ``it is easier for me to read text list.'' The \textit{Hierarchical List} was deemed useful to start the analysis from \textit{leading conditions}, that is, conditions that appear first in a rule. Several participants found it easier to start from this view. For example, P2 used this view to answer the question ``what are the conditions that by their own lead to a quite pure node?'' While answering this type of questions is possible in the feature-aligned tree, some participants found it useful to find all the child conditions of a leading condition clustered in the same spatial area. P6 commented that ``\textit{the hierarchical rule list is useful to look at a group of rules with the same leading condition}''. The \textit{Feature-Aligned Tree} was used to explore the widest variety of tasks. P1 and P3 mentioned that they used the feature-aligned tree to check how the nodes becomes more and more pure (in terms of the model prediction) along the decision path. P3 liked the squared node in this view because ``\textit{(they) are good at comparing errors (by class)}''. P6 commented that ``\textit{(the feature-aligned tree) is good at finding the rules (that) contain different values of a feature}'', which is what we expected by aligning features in the same column. Although some experts used hierarchical list to \textit{search rules} for analysis, the experts also searched the nodes with desired properties in this view to analyze the model behavior of interest. For example, P2 and P7 looked for the largest node first and then extracted the \textit{feature-outcome logic}. P7 also liked the column ordering in this view because ``it helped people focus on important features.'' \looseness=-1


\textbf{Reasoning driven by short rules.} Another surprising finding is that our participants almost never reasoned about more than two conditions at a time. While {\sure} can generate rules of arbitrary length and the visualizations inspected by the participants included rules longer than two, most of the reasoning we observed included maximum two conditions. Most commonly, our participants reasoned extensively about the effect of \textit{individual} features, looking for effects of individual conditions on outcomes and gauging the relevance of features in general and for specific subgroups. When participants reasoned about more complex relationships they almost always involved maximum two features. 
\looseness=-1

\textbf{User-driven hypothesis testing and exploration.} During the analysis sessions it was rather common to observe the participants wanting to go beyond the set of rules extracted by the algorithm. This necessity is exemplified by the frequency with which participants wanted to set their own thresholds to see the effect on the outcome. We did not anticipate how often users would voice their desire to build their own individual rules to test hypotheses they had and were not included in the result set. 


\textbf{Effect of domain expertise.} We noticed a marked difference between the way participants with a stronger domain knowledge approached the problem and used the system in comparison to the participants who were not familiar with the domain problem. Users with domain expertise tended to generate a lot of their questions from intuitions and hunches they had and to use the tool to verify whether their intuition matched the patterns and logic captured by the rules. Users with less or no domain knowledge used {\sure} in a much more exploratory fashion, with much fewer questions generated by their own intuitions.

\section{Discussion}
As shown in the studies, the feature-aligned tree visualization worked surprisingly well. Users were able to perform nontrivial \textit{visual interpretation} and \textit{visual search} tasks with high accuracy. Subjectively, participants also felt that carrying out the tasks with the aligned trees was rather easy. Considering that the participants used this visualization technique for the first time and that they had very minimal training, the results are encouraging and suggest that complex link-node visualizations like the aligned-tree could be used proficiently in the context of rule visualization. The results are also corroborated by our observations in the observational study, during which most participants were able to perform complex inferential tasks (\textit{Search / Explore} and \textit{Explain}) using the tree visualization.


Another surprising finding is how frequently the participants switched between the tree visualization and the \textit{text list} and the \textit{hierarchical list}. Even though we initially introduced these visualizations to help users smoothly transition from more familiar views to the tree visualization, we found the three visualizations to provide complementary support for the users. The participants performed different analytical tasks with different visualizations; often switching back and forth between the different techniques. In turn, this raises the question of how to design visual analytics systems that provide complementary views. While many experimental studies tend to focus on defining which view is ``best'', it seems reasonable to investigate in more depth how different visualization techniques can be used in a \textit{complementary} fashion to support a variety of tasks for which they perform synergistically. This is certainly true in rule visualization, where the existing research on how to visualize rules and their conditions is particularly limited~\cite{yuan2021exploration}.
\looseness=-1

Despite obtaining positive results in the evaluations, it is important to keep in mind that the context of these studies is limited to the specific data sets and models we used. One important limitation is the focus on binary classification. While we have anecdotal evidence that {\sure} can visualize data sets with 3-5 classes, we did not investigate in sufficient depth how to make it visually scale to data sets with a much larger number of classes. We expect that the visual designs we proposed would need some adjustments or ways to reduce the complexity. Similarly, while we know {\sure} can easily handle data sets with a high number of features, it seems plausible to believe specific feature reduction mechanisms will be needed for very high-dimensional data sets. 

As shown in the experiments on benchmark data sets, the algorithm we introduced can handle different user-defined parameters to reach the goal of \textit{completeness} and \textit{parsimony} in a reasonable amount of time. Similar to the visual scalability problem, the algorithm has better and more stable performance on binary rather than multi-class classification. More research is needed to further improve performance on multiple class data. These experiments also demonstrate the trade-off between low complexity and high faithfulness previously mentioned. In this respect the results emphasize the importance of user-driven parameterization for the kind of systems proposed in this paper. 


The task characterization identified interesting trends and gaps. A relevant observation is that most tasks were in the three categories \textit{Rule Search / Exploration}, \textit{Feature-Outcome Logic}, and \textit{Outcome-Feature Logic}. Designers of systems that use rules to explain a machine learning model can start from this characterization to support a similar set of tasks. Starting from the table we provided, system designers can reason about what tasks they may need to support and which to cover with their design. The same characterization can also be used for evaluation purposes to ideate tasks to use in a user study. As mentioned before, the characterization also helped us identify relevant gaps; and above all the need to support users in defining and exploring their own manually generated rules. Since we did not anticipate this need, {\sure} was not designed to support this type of task. Future work will be needed to address this need and explore it further. In particular, future research is needed to better understand how different visual analytics paradigms support different needs. More specifically, systems like {\sure} start from the assumption that interaction will be needed exclusively ``downstream'' to explore and interpret the results of the automated model building procedure (a set of rules in this case). However, such systems can also be designed starting from the assumptions that users want to be able to build their own models manually or semi-automatically, or at the very least be able to modify their output. How to do this effectively is an open challenge for visualization research that needs to be addressed in future work (Tam \textit{et al.} provide interesting initial research in this direction~\cite{tam2016analysis}).

Another relevant observation from the study is that participants with a strong background knowledge tended to use {\sure} in a different way than those who did not have domain knowledge. Domain experts tended to start way more often from their intuitions and used {\sure} more to \textit{verify} them then to explore the model's behavior freely. Participants who lacked background knowledge were way less focused and tended to be guided by what stood out in the visualization. An interesting methodological implication of this observation is that evaluating systems excursively with users who do not have a strong domain knowledge can greatly skew and limit the results of a study. \looseness=-1

The observation that domain experts only explored maximum one or two features at once has potentially very strong implications on how rule-based systems should be designed. If reasoning with more than two (or maybe three) features turns out to be an inherently difficult task and a fundamental limitation in human reasoning, designing systems that falsely expect users to achieve more than what is possible may be greatly limited. While it is not known whether this is true or not, it is crucial to investigate this aspect further. In particular, it is necessary to study in more depth how people reason with rules and what people are capable of doing with rules. Once such research is available, it will be easier to design solutions that keep cognitive limitations in mind. For example, the results of this research could be used to decide whether it is better to privilege rule sets with many short rules or rule sets with fewer but longer rules. \looseness=-1

\section{Conclusion}
In this work, we presented an interactive workflow, {\sure}, that integrates rule generation, rule visualization and exploration. We also introduced the idea of visualizing rules with a hierarchical structure through a novel visualization technique called feature-aligned trees. We presented experiments that evaluated the proposed rule generation algorithm, the novel visualization techniques and the visual analytics system. The studies helped verify where the introduced solutions produced satisfactory results and where interesting gaps existed. The user studies also helped develop several interesting observations that can be useful to inspire future research for the advancement of understanding how to design effective rule-based visual analytics solutions.
\looseness=-1


%





\ifCLASSOPTIONcaptionsoff
  \newpage
\fi



\bibliographystyle{IEEEtran}
\bibliography{ref}

\begin{thebibliography}{10}
\providecommand{\url}[1]{#1}
\csname url@samestyle\endcsname
\providecommand{\newblock}{\relax}
\providecommand{\bibinfo}[2]{#2}
\providecommand{\BIBentrySTDinterwordspacing}{\spaceskip=0pt\relax}
\providecommand{\BIBentryALTinterwordstretchfactor}{4}
\providecommand{\BIBentryALTinterwordspacing}{\spaceskip=\fontdimen2\font plus
\BIBentryALTinterwordstretchfactor\fontdimen3\font minus
  \fontdimen4\font\relax}
\providecommand{\BIBforeignlanguage}[2]{{%
\expandafter\ifx\csname l@#1\endcsname\relax
\typeout{** WARNING: IEEEtran.bst: No hyphenation pattern has been}%
\typeout{** loaded for the language `#1'. Using the pattern for}%
\typeout{** the default language instead.}%
\else
\language=\csname l@#1\endcsname
\fi
#2}}
\providecommand{\BIBdecl}{\relax}
\BIBdecl

\bibitem{hastie1990generalized}
T.~J. Hastie and R.~J. Tibshirani, \emph{Generalized additive models}.\hskip
  1em plus 0.5em minus 0.4em\relax CRC press, 1990, vol.~43.

\bibitem{safavian1991survey}
S.~R. Safavian and D.~Landgrebe, ``A survey of decision tree classifier
  methodology,'' \emph{IEEE transactions on systems, man, and cybernetics},
  vol.~21, no.~3, pp. 660--674, 1991.

\bibitem{frank1998generating}
E.~Frank and I.~H. Witten, ``Generating accurate rule sets without global
  optimization,'' 1998.

\bibitem{hastie2017generalized}
T.~J. Hastie and R.~J. Tibshirani, \emph{Generalized additive models}.\hskip
  1em plus 0.5em minus 0.4em\relax Routledge, 2017.

\bibitem{hohman2019gamut}
F.~Hohman, A.~Head, R.~Caruana, R.~DeLine, and S.~M. Drucker, ``Gamut: A design
  probe to understand how data scientists understand machine learning models,''
  in \emph{Proceedings of the 2019 CHI conference on human factors in computing
  systems}, 2019, pp. 1--13.

\bibitem{molnar2019}
C.~Molnar, \emph{Interpretable Machine Learning}, 2019,
  \url{https://christophm.github.io/interpretable-ml-book/}.

\bibitem{agrawal1994fast}
R.~Agrawal, R.~Srikant \emph{et~al.}, ``Fast algorithms for mining association
  rules,'' in \emph{Proc. 20th int. conf. very large data bases, VLDB}, vol.
  1215, 1994, pp. 487--499.

\bibitem{flach2001confirmation}
P.~A. Flach and N.~Lachiche, ``Confirmation-guided discovery of first-order
  rules with tertius,'' \emph{Machine learning}, vol.~42, no. 1-2, pp. 61--95,
  2001.

\bibitem{pei2000closet}
J.~Pei, J.~Han, R.~Mao \emph{et~al.}, ``Closet: An efficient algorithm for
  mining frequent closed itemsets.'' in \emph{ACM SIGMOD workshop on research
  issues in data mining and knowledge discovery}, vol.~4, no.~2, 2000, pp.
  21--30.

\bibitem{lakkaraju2019faithful}
H.~Lakkaraju, E.~Kamar, R.~Caruana, and J.~Leskovec, ``Faithful and
  customizable explanations of black box models,'' in \emph{Proceedings of the
  2019 AAAI/ACM Conference on AI, Ethics, and Society}, 2019, pp. 131--138.

\bibitem{rajapaksha2020lormika}
D.~Rajapaksha, C.~Bergmeir, and W.~Buntine, ``Lormika: Local rule-based model
  interpretability with k-optimal associations,'' \emph{Information Sciences},
  vol. 540, pp. 221--241, 2020.

\bibitem{moradi2021post}
M.~Moradi and M.~Samwald, ``Post-hoc explanation of black-box classifiers using
  confident itemsets,'' \emph{Expert Systems with Applications}, vol. 165, p.
  113941, 2021.

\bibitem{pal2001fuzzy}
N.~R. Pal and S.~Chakraborty, ``Fuzzy rule extraction from id3-type decision
  trees for real data,'' \emph{IEEE Transactions on Systems, Man, and
  Cybernetics, Part B (Cybernetics)}, vol.~31, no.~5, pp. 745--754, 2001.

\bibitem{quinlan1987generating}
J.~R. Quinlan, ``Generating production rules from decision trees.'' in
  \emph{ijcai}, vol.~87.\hskip 1em plus 0.5em minus 0.4em\relax Citeseer, 1987,
  pp. 304--307.

\bibitem{craven1996extracting}
M.~Craven and J.~W. Shavlik, ``Extracting tree-structured representations of
  trained networks,'' in \emph{Advances in neural information processing
  systems}, 1996, pp. 24--30.

\bibitem{augasta2012reverse}
M.~G. Augasta and T.~Kathirvalavakumar, ``Reverse engineering the neural
  networks for rule extraction in classification problems,'' \emph{Neural
  processing letters}, vol.~35, no.~2, pp. 131--150, 2012.

\bibitem{zilke2016deepred}
J.~R. Zilke, E.~L. Menc{\'\i}a, and F.~Janssen, ``Deepred--rule extraction from
  deep neural networks,'' in \emph{International Conference on Discovery
  Science}.\hskip 1em plus 0.5em minus 0.4em\relax Springer, 2016, pp.
  457--473.

\bibitem{bastani2017interpreting}
O.~Bastani, C.~Kim, and H.~Bastani, ``Interpreting blackbox models via model
  extraction,'' \emph{arXiv preprint arXiv:1705.08504}, 2017.

\bibitem{thiagarajan2016treeview}
J.~J. Thiagarajan, B.~Kailkhura, P.~Sattigeri, and K.~N. Ramamurthy,
  ``Treeview: Peeking into deep neural networks via feature-space
  partitioning,'' \emph{arXiv preprint arXiv:1611.07429}, 2016.

\bibitem{velmurugan2021evaluating}
M.~Velmurugan, C.~Ouyang, C.~Moreira, and R.~Sindhgatta, ``Evaluating fidelity
  of explainable methods for predictive process analytics,'' in
  \emph{International Conference on Advanced Information Systems
  Engineering}.\hskip 1em plus 0.5em minus 0.4em\relax Springer, 2021, pp.
  64--72.

\bibitem{cameron1997r}
A.~C. Cameron and F.~A. Windmeijer, ``An r-squared measure of goodness of fit
  for some common nonlinear regression models,'' \emph{Journal of
  econometrics}, vol.~77, no.~2, pp. 329--342, 1997.

\bibitem{messalas2019model}
A.~Messalas, Y.~Kanellopoulos, and C.~Makris, ``Model-agnostic interpretability
  with shapley values,'' in \emph{2019 10th International Conference on
  Information, Intelligence, Systems and Applications (IISA)}.\hskip 1em plus
  0.5em minus 0.4em\relax IEEE, 2019, pp. 1--7.

\bibitem{lundberg2017unified}
S.~M. Lundberg and S.-I. Lee, ``A unified approach to interpreting model
  predictions,'' in \emph{Advances in neural information processing systems},
  2017, pp. 4765--4774.

\bibitem{poursabzi2021manipulating}
F.~Poursabzi-Sangdeh, D.~G. Goldstein, J.~M. Hofman, J.~W. Wortman~Vaughan, and
  H.~Wallach, ``Manipulating and measuring model interpretability,'' in
  \emph{Proceedings of the 2021 CHI Conference on Human Factors in Computing
  Systems}, 2021, pp. 1--52.

\bibitem{lakkaraju2016interpretable}
H.~Lakkaraju, S.~H. Bach, and J.~Leskovec, ``Interpretable decision sets: A
  joint framework for description and prediction,'' in \emph{Proceedings of the
  22nd ACM SIGKDD international conference on knowledge discovery and data
  mining}, 2016, pp. 1675--1684.

\bibitem{wang2017bayesian}
T.~Wang, C.~Rudin, F.~Doshi-Velez, Y.~Liu, E.~Klampfl, and P.~MacNeille, ``A
  bayesian framework for learning rule sets for interpretable classification,''
  \emph{The Journal of Machine Learning Research}, vol.~18, no.~1, pp.
  2357--2393, 2017.

\bibitem{lage2019evaluation}
I.~Lage, E.~Chen, J.~He, M.~Narayanan, B.~Kim, S.~Gershman, and F.~Doshi-Velez,
  ``An evaluation of the human-interpretability of explanation,'' \emph{arXiv
  preprint arXiv:1902.00006}, 2019.

\bibitem{lakkaraju2017interpretable}
H.~Lakkaraju, E.~Kamar, R.~Caruana, and J.~Leskovec, ``Interpretable \&
  explorable approximations of black box models,'' \emph{arXiv preprint
  arXiv:1707.01154}, 2017.

\bibitem{guidotti2018local}
R.~Guidotti, A.~Monreale, S.~Ruggieri, D.~Pedreschi, F.~Turini, and
  F.~Giannotti, ``Local rule-based explanations of black box decision
  systems,'' \emph{arXiv preprint arXiv:1805.10820}, 2018.

\bibitem{ribeiro2018anchors}
M.~T. Ribeiro, S.~Singh, and C.~Guestrin, ``Anchors: High-precision
  model-agnostic explanations,'' in \emph{Proceedings of the AAAI conference on
  artificial intelligence}, vol.~32, no.~1, 2018.

\bibitem{wang2015falling}
F.~Wang and C.~Rudin, ``Falling rule lists,'' in \emph{Artificial Intelligence
  and Statistics}, 2015, pp. 1013--1022.

\bibitem{schulz2011treevis}
H.-J. Schulz, ``Treevis.net: A tree visualization reference,'' \emph{IEEE
  Computer Graphics and Applications}, vol.~31, no.~6, pp. 11--15, 2011.

\bibitem{van2011baobabview}
S.~Van Den~Elzen and J.~J. Van~Wijk, ``Baobabview: Interactive construction and
  analysis of decision trees,'' in \emph{2011 IEEE conference on visual
  analytics science and technology (VAST)}.\hskip 1em plus 0.5em minus
  0.4em\relax IEEE, 2011, pp. 151--160.

\bibitem{zarlenga2021efficient}
M.~E. Zarlenga, Z.~Shams, and M.~Jamnik, ``Efficient decompositional rule
  extraction for deep neural networks,'' \emph{arXiv preprint
  arXiv:2111.12628}, 2021.

\bibitem{yuan2021exploration}
J.~Yuan, O.~Nov, and E.~Bertini, ``An exploration and validation of visual
  factors in understanding classification rule sets,'' \emph{arXiv preprint
  arXiv:2109.09160}, 2021.

\bibitem{letham2015interpretable}
B.~Letham, C.~Rudin, T.~H. McCormick, D.~Madigan \emph{et~al.}, ``Interpretable
  classifiers using rules and bayesian analysis: Building a better stroke
  prediction model,'' \emph{The Annals of Applied Statistics}, vol.~9, no.~3,
  pp. 1350--1371, 2015.

\bibitem{subramanian1992comparison}
G.~H. Subramanian, J.~Nosek, S.~P. Raghunathan, and S.~S. Kanitkar, ``A
  comparison of the decision table and tree,'' \emph{Communications of the
  ACM}, vol.~35, no.~1, pp. 89--94, 1992.

\bibitem{vessey1986structured}
I.~Vessey and R.~Weber, ``Structured tools and conditional logic: An empirical
  investigation,'' \emph{Communications of the ACM}, vol.~29, no.~1, pp.
  48--57, 1986.

\bibitem{huysmans2011empirical}
J.~Huysmans, K.~Dejaeger, C.~Mues, J.~Vanthienen, and B.~Baesens, ``An
  empirical evaluation of the comprehensibility of decision table, tree and
  rule based predictive models,'' \emph{Decision Support Systems}, vol.~51,
  no.~1, pp. 141--154, 2011.

\bibitem{ming2019rulematrix}
Y.~Ming, H.~Qu, and E.~Bertini, ``Rulematrix: Visualizing and understanding
  classifiers with rules,'' \emph{IEEE transactions on visualization and
  computer graphics}, vol.~25, no.~1, pp. 342--352, 2019.

\bibitem{di2019surrogate}
F.~Di~Castro and E.~Bertini, ``Surrogate decision tree visualization,'' 2019.

\bibitem{neto2020explainable}
M.~P. Neto and F.~V. Paulovich, ``Explainable matrix--visualization for global
  and local interpretability of random forest classification ensembles,''
  \emph{arXiv preprint arXiv:2005.04289}, 2020.

\bibitem{estivill2020human}
V.~Estivill-Castro, E.~Gilmore, and R.~Hexel, ``Human-in-the-loop construction
  of decision tree classifiers with parallel coordinates,'' in \emph{2020 IEEE
  International Conference on Systems, Man, and Cybernetics (SMC)}.\hskip 1em
  plus 0.5em minus 0.4em\relax IEEE, 2020, pp. 3852--3859.

\bibitem{zhao2019iforest}
X.~Zhao, Y.~Wu, D.~L. Lee, and W.~Cui, ``iforest: Interpreting random forests
  via visual analytics,'' \emph{IEEE transactions on visualization and computer
  graphics}, vol.~25, no.~1, pp. 407--416, 2019.

\bibitem{asuncion2007uci}
A.~Asuncion and D.~Newman, ``Uci machine learning repository,'' 2007.

\bibitem{romano2021pmlb}
J.~D. Romano, T.~T. Le, W.~La~Cava, J.~T. Gregg, D.~J. Goldberg,
  P.~Chakraborty, N.~L. Ray, D.~Himmelstein, W.~Fu, and J.~H. Moore, ``Pmlb
  v1.0: an open source dataset collection for benchmarking machine learning
  methods,'' \emph{arXiv preprint arXiv:2012.00058v2}, 2021.

\bibitem{smith1988using}
J.~W. Smith, J.~E. Everhart, W.~Dickson, W.~C. Knowler, and R.~S. Johannes,
  ``Using the adap learning algorithm to forecast the onset of diabetes
  mellitus,'' in \emph{Proceedings of the annual symposium on computer
  application in medical care}.\hskip 1em plus 0.5em minus 0.4em\relax American
  Medical Informatics Association, 1988, p. 261.

\bibitem{fico}
FICO, ``Explainable machine learning challenge,''
  \url{https://community.fico.com/s/explainable-machine-learning-challenge?tabset-3158a=2},
  2018.

\bibitem{fonteyn1993description}
M.~E. Fonteyn, B.~Kuipers, and S.~J. Grobe, ``A description of think aloud
  method and protocol analysis,'' \emph{Qualitative health research}, vol.~3,
  no.~4, pp. 430--441, 1993.

\bibitem{tam2016analysis}
G.~K. Tam, V.~Kothari, and M.~Chen, ``An analysis of machine-and
  human-analytics in classification,'' \emph{IEEE transactions on visualization
  and computer graphics}, vol.~23, no.~1, pp. 71--80, 2016.

\end{thebibliography}
%




%

\begin{IEEEbiography}[{\includegraphics[width=1in,height=1.25in,clip,keepaspectratio]{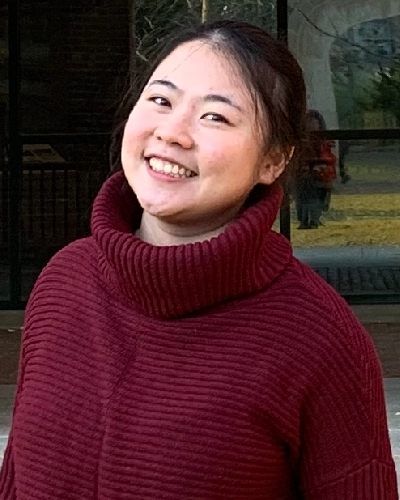}}]{Jun Yuan} is currently a Ph.D. candidate in the Department of Computer Science and Engineering at New York University. She obtained a B.S. in Software Engineering from Fudan University, China in 2017. Her research interest lies in the intersection of Data Visualization, Human-Computer Interaction and Explainable AI (xAI). For more information see https://junyuanjun.github.io.
\end{IEEEbiography}

\begin{IEEEbiography}[{\includegraphics[width=1in,height=1.25in,clip,keepaspectratio]{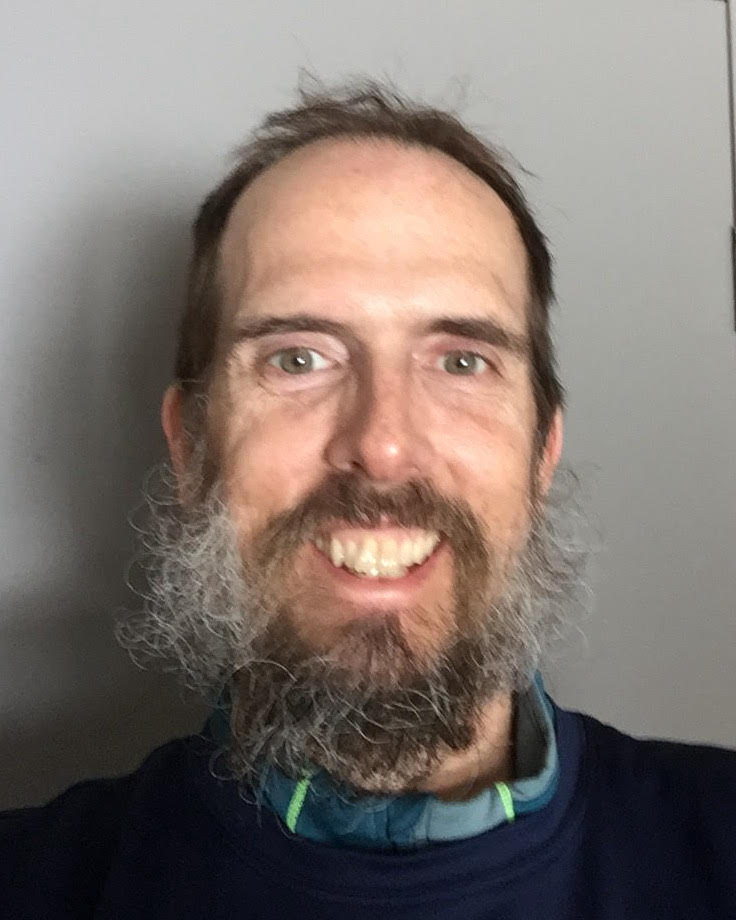}}]{Brian Barr} is an explainable artificial intelligence researcher at Capital One. His research interests include deep learning on tabular data, representation learning, and local attribution methods. He received a Ph.D. in Mechanical Engineering from Clarkson University.
\end{IEEEbiography}

\begin{IEEEbiography}[{\includegraphics[width=1in,height=1.25in,clip,keepaspectratio]{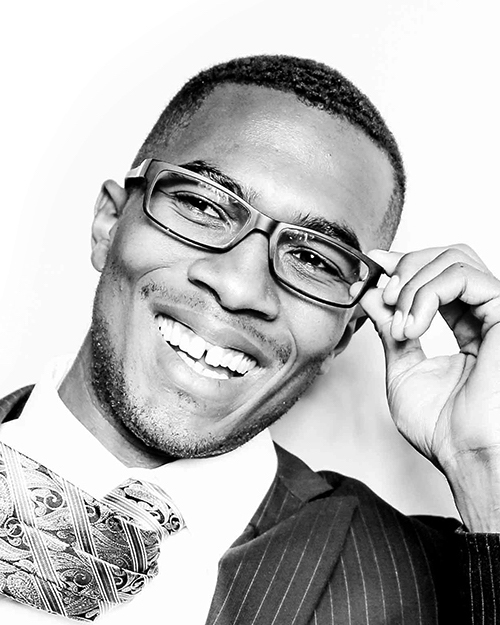}}]{Kyle Overton} is a Visual/UX designer at Capital One. He is passionate for creating designs that embody a sense of awe, while delivering powerful messages of personal well-being, social change, and environmental care. He received a M.S. in Informatics - Human-Computer Interaction/design from Indiana University Bloomington.

\end{IEEEbiography}

\begin{IEEEbiography}[{\includegraphics[width=1in,height=1.25in,clip,keepaspectratio]{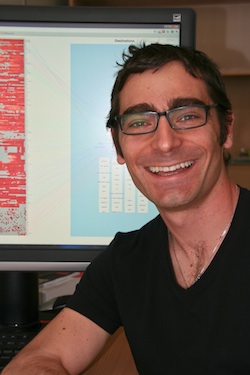}}]{Enrico Bertini} is an Associate Professor in the Department of Computer Science and Engineering at NYU Tandon School of Engineering. His research focuses on the development and evaluation of interactive visual interfaces to help scientists, researchers and domain experts, reason with data and machine learning models. He is also the host of Data Stories, a popular podcast on data visualization.

\end{IEEEbiography}







\end{document}